\documentclass[apjl,iop]{emulateapj}
\usepackage{comment}
\usepackage{ifthen}
\usepackage{ textcomp }

\newcommand{\forloop}[5][1]%
{%
\setcounter{#2}{#3}%
\ifthenelse{#4}%
	{%
	#5%
	\addtocounter{#2}{#1}%
	\forloop[#1]{#2}{\value{#2}}{#4}{#5}%
	}%
	{%
	}%
}%

\newcommand{\vsini}{\ensuremath{v \sin{i}}}
\newcommand{\feh}{\ensuremath{\rm [Fe/H]}}

\newcommand{\vmac}{\ensuremath{v_{\rm mac}}}
\newcommand{\vmic}{\ensuremath{v_{\rm mic}}}

\newcommand{\masy}{\ensuremath{\rm mas\,yr^{-1}}}

\newcommand{\gcmc}{\ensuremath{\rm g\,cm^{-3}}}

\newcommand{\rsun}{\ensuremath{R_\sun}}
\newcommand{\msun}{\ensuremath{M_\sun}}
\newcommand{\lsun}{\ensuremath{L_\sun}}

\newcommand{\rstar}{\ensuremath{R_\star}}
\newcommand{\mstar}{\ensuremath{M_\star}}
\newcommand{\lstar}{\ensuremath{L_\star}}

\newcommand{\teffstar}{\ensuremath{T_{\rm eff\star}}}
\newcommand{\rhostar}{\ensuremath{\rho_\star}}
\newcommand{\loggstar}{\ensuremath{\log{g_{\star}}}}

\newcommand{\rpl}{\ensuremath{R_{p}}}
\newcommand{\mpl}{\ensuremath{M_{p}}}

\newcommand{\rhopl}{\ensuremath{\rho_{p}}}

\newcommand{\arstar}{\ensuremath{a/\rstar}}
\newcommand{\zrstar}{\ensuremath{\zeta/\rstar}}

\newcommand{\rjup}{\ensuremath{R_{\rm J}}}
\newcommand{\mjup}{\ensuremath{M_{\rm J}}}

\newcommand{\reffig}[1]{Fig.~\ref{fig:#1}}
\newcommand{\refsec}[1]{\mbox{\S\ \ref{sec:#1}}}

\newcommand{\reffigl}[1]{Figure~\ref{fig:#1}}
\newcommand{\refsecl}[1]{\mbox{Section \ref{sec:#1}}}

\newcommand{\reftabl}[1]{Table~\ref{tab:#1}}

\newcommand{\loopand}{\ifnum\value{planetcounter}=2 and \else\fi}
\newcommand{\loopcomma}{\ifnum\value{planetcounter}<2 ,\else. \fi}
\newcommand{\loopcommanoperiod}{\ifnum\value{planetcounter}<2 ,\else \space\fi}
\newcommand{\loopcommanospace}{\ifnum\value{planetcounter}<2 ,\else \fi}

\newcounter{planetcounter}

\newboolean{emulateapj}
\setboolean{emulateapj}{true}

\newboolean{rvtablelong}
\setboolean{rvtablelong}{true}

\newboolean{astroph}
\setboolean{astroph}{true}

\shortauthors{Ciceri et al.}
\shorttitle{HATS-15\lowercase{b} and HATS-16\lowercase{b}}
\ifthenelse{\boolean{emulateapj}}{
    
}{
    
}

\begin{document}

\title{
HATS-15\lowercase{b} and HATS-16\lowercase{b}: Two massive planets transiting old G dwarf stars
}

\author{Ciceri, S.$^{1}$, 
Mancini, L.$^{1}$,
Henning, T.$^{1}$,
Bakos, G.$^{2}$,
Penev, K.$^{2}$,
Brahm, R.$^{3,4}$,
Zhou, G.$^{5}$,
Hartman, J. D.$^{2}$,
Bayliss, D.$^{6}$,
Jord\'an, A.$^{3,4}$,
Csubry, Z.$^{2}$,
de Val-Borro, M.$^{2}$,
Bhatti, W.$^{2}$,
Rabus, M.$^{3,1}$,
Espinoza, N.$^{3,4}$,
Suc, V.$^{3}$,
Schmidt, B.$^{5}$,
Noyes, R.$^{7}$,
Howard, A. W.$^{8}$,
Fulton, B. J.$^{8}$,
Isaacson, H.$^{9}$,
Marcy, G. W.$^{9}$,
Butler, R. P.$^{10}$,
Arriagada, P.$^{10}$,
Crane, J.$^{10}$,
Shectman, S.$^{10}$,
Thompson, I.$^{10}$,
Tan, T. G.$^{11}$,
L\'az\'ar, J.$^{12}$,
Papp, I.$^{12}$,
Sari, P.$^{12}$\\
}

\altaffiltext{}{
$^{1}$ Max Planck Institute for Astronomy, K\"{o}nigstuhl 17, 69117, Heidelberg, Germany \\
$^{2}$ Department of Astrophysical Sciences, Princeton University, Princeton, NJ 08544, USA\\
$^{3}$ Instituto de Astrof\'{i}sica, Pontificia Universidad Cat\'{o}lica de Chile, Av. Vicu\~{n}a Mackenna 4860, 7820436 Macul, Santiago, Chile\\ %
$^{4}$ Millennium Institute of Astrophysics, Av. Vicu\~{n}a Mackenna 4860, 7820436 Macul, Santiago, Chile\\ %
$^{5}$ The Australian National University, Canberra, Australia\\
$^{6}$ Observatoire Astronomique de l'Universit\'{e} de Gen\`{e}ve, 51 ch. des Maillettes, 1290 Versoix, Switzerland\\
$^{7}$ Harvard-Smithsonian Center for Astrophysics, Cambridge, MA 02138 USA\\ %
$^{8}$ Institute for Astronomy, University of Hawaii at Manoa,Honolulu, HI, USA\\ %
$^{9}$ Department of Astronomy, University of California, Berkeley, CA 94720-3411, USA\\
$^{10}$ Carnegie Institution of Washington Department of Terrestrial Magnetism, NW Washington, DC 20015-1305, USA\\
$^{11}$ Perth Exoplanet Survey Telescope, Perth, Australia\\
$^{12}$ Hungarian Astronomical Association, Budapest, Hungary\\ %
The HATSouth network is operated by a collaboration consisting of Princeton University (PU), the Max Planck Institute f\"ur Astronomie (MPIA), the Australian National University (ANU), and the Pontificia Universidad Cat\'olica de Chile (PUC).  The station at Las Campanas Observatory (LCO) of the Carnegie Institute is operated by PU in conjunction with PUC, the station at the High Energy Spectroscopic Survey (H.E.S.S.) site is operated in conjunction with MPIA, and the station at Siding Spring Observatory (SSO) is operated jointly with ANU.
Based in part on observations performed at the ESO La Silla Observatory in Chile, with the Coralie and FEROS spectrographs mounted on the Euler-Swiss and  MPG~2.2\,m telescopes respectively. This paper includes data gathered with the 6.5 m Magellan Telescopes located as Las Campanas Observatory, Chile. 
Based in part on data collected at Keck Telescope. Observations obtained with facilities of the Las
Cumbres Observatory Global Telescope are used in this paper.
}

\begin{abstract}

\setcounter{footnote}{16}
We report the discovery of HATS-15\,b and HATS-16\,b, two massive transiting extrasolar planets orbiting evolved ($\sim 10$\,Gyr) main-sequence stars. The planet HATS-15\,b, which is hosted by a G9\,V star ($V=14.8$\,mag), is a hot Jupiter with  mass of $2.17\pm0.15\, M_{\mathrm{J}}$ and radius of $1.105\pm0.0.040\, R_{\mathrm{J}}$, and completes its orbit in nearly 1.7 days. HATS-16\,b is a very massive hot Jupiter with mass of $3.27\pm0.19\, M_{\mathrm{J}}$ and radius of  $1.30\pm0.15\, R_{\mathrm{J}}$; it orbits around its G3\,V parent star ($V=13.8$\,mag) in $\sim2.7$\,days. HATS-16 is slightly active and shows a periodic photometric modulation, implying a rotational period of 12 days which is unexpectedly short given its isochronal age. This fast rotation might be the result of the tidal interaction between the star and its planet.

\setcounter{footnote}{0}
\end{abstract}
\keywords{
    planetary systems ---
    stars: individual (
\setcounter{planetcounter}{1}
HATS-15\loopcommanoperiod
\setcounter{planetcounter}{2}
HATS-16, GSC 7516-00867 \loopcommanoperiod
\setcounter{planetcounter}{3}
) 
 --  techniques: spectroscopic, photometric
}


\section{Introduction}
\label{sec:introduction}

Before the {\it Kepler} mission \citep{borucki:2010}, most of the exoplanets discovered with the transit method by the ground-based surveys \citep[e.g. WASP and HATNet,][]{pollacco:2006, bakos:2004:hatnet} were hot Jupiters (i.e. those planets in the Jupiter-mass regime that circle very close-in their host star).
This is essentially due to the relative ease with which this kind of exoplanet can be detected: indeed both the radial velocity (RV) and transit detection techniques are biased towards finding massive planets at short periods. Nowadays, with an increased sample of more than 1800 planets, we realize that the hot-Jupiter occurrence is just a tiny fraction, $\lesssim 1\%$ for solar-like stars, compared to the larger number of smaller (Neptunian and rocky) planets \citep[e.g.][]{mayor:2011,howard:2012,dong:2013,fressin:2013}.

However, hot-Jupiters are still of interest to astrophysicists because many of their properties are not well understood. In particular, it is not clear what are the physical mechanisms that cause them to migrate from their formation region down to $\sim 10^{-2}$\,AU from the parent stars; it is also very puzzling that many of the known hot Jupiters have a radius larger than what predicted by standard models of structure of gaseous giant planets; hot Jupiters with masses $>2\,M_{\mathrm{J}}$ are, on the other hand, more an exception than a rule \citep[e.g.][]{jiang:2007}. In this work, we present two new hot-Jupiters transiting planets, HATS-15b and HATS-16b, belonging to the class of massive gas planets.

The HATS-15 and HATS-16 planetary systems have been discovered within the HATSouth ground-based survey \citep[e.g.][]{bakos:2013:hatsouth,penev:2013:hats1}. HATSouth is a network of six completely automated units (named HS-1 to HS-6), which are stationed in pairs at three different sites in the southern hemisphere: Las Campanas Observatory in Chile (LCO), the High Energy Spectroscopic System (HESS) site in Namibia and the Siding Spring Observatory (SSO) in Australia. The main part of each station is the mount on which four 18\,cm astrographs and four CCD cameras are lodged. The mutual distance of the three sites, nearly $120^{\circ}$ from each other, permits operations of at least one station at any time, allowing a continuous 24--hour monitoring of a stellar field. 

\ifthenelse{\boolean{emulateapj}}{
    \begin{figure*}[!ht]
}{
    \begin{figure}[!ht]
}
\plottwo{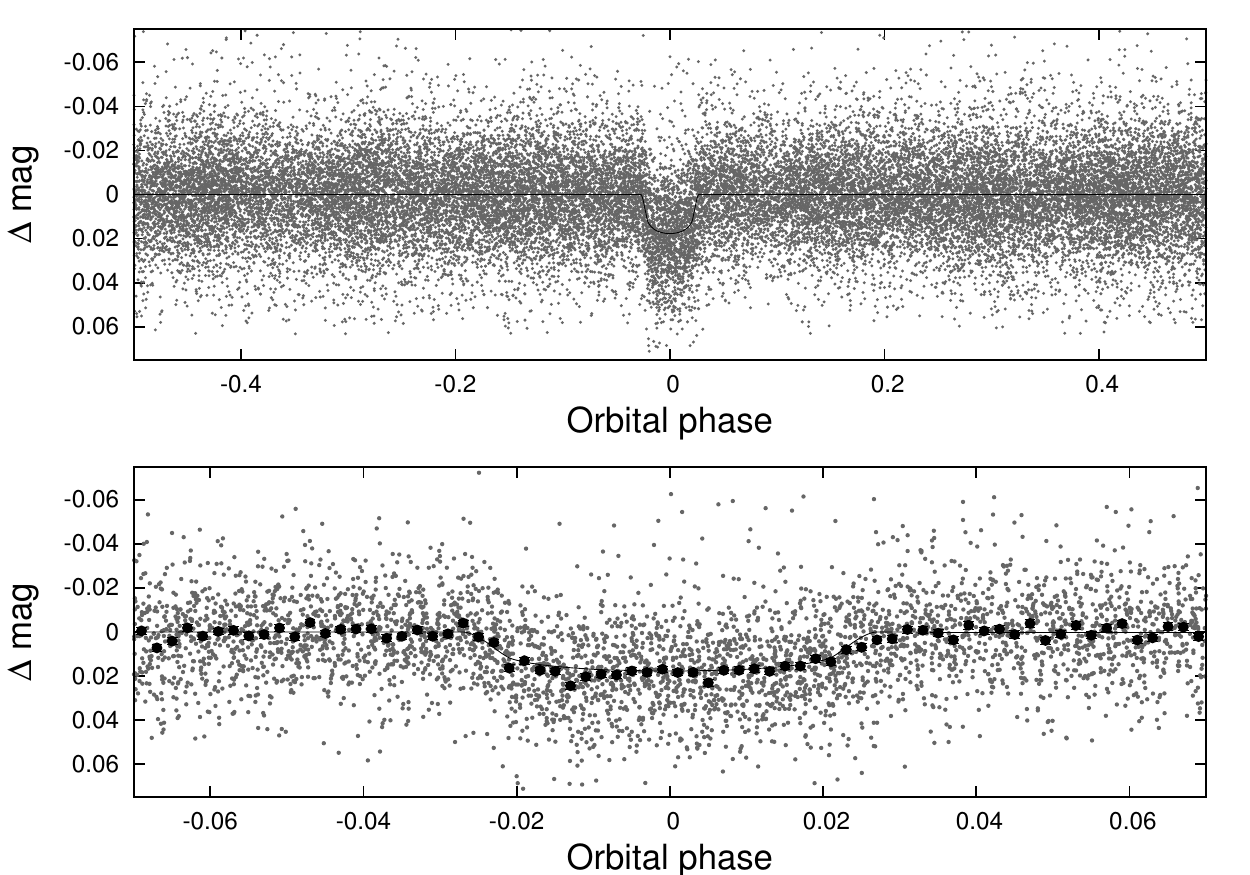}{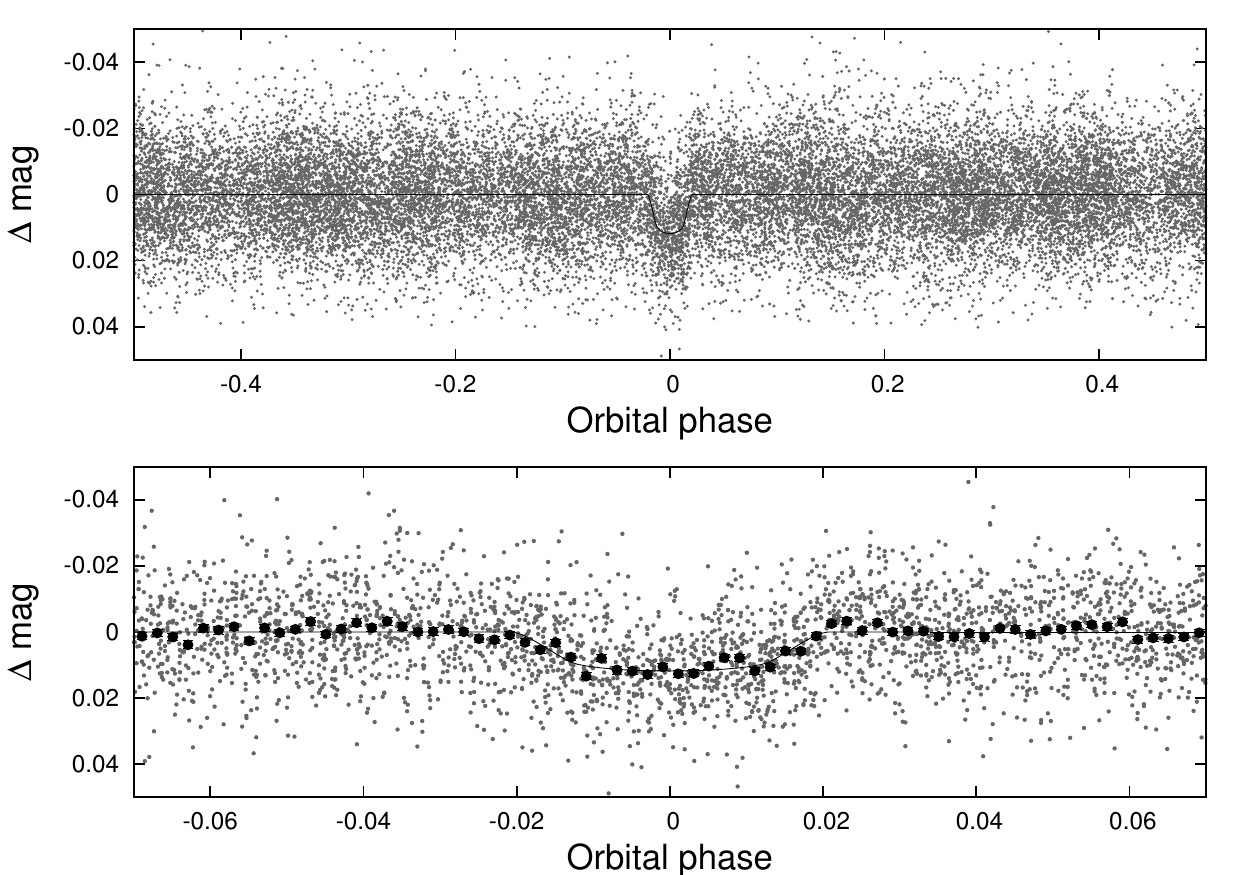}
\caption[]{
HATSouth photometry for HATS-15 (left) and HATS-16 (right). The top panel of each system shows the relative full light curve phase-folded and unbinned; the superimposed solid lines are the model fits to the light curves. The bottom panels show the light curve zoomed-in on the transit. The dark filled circles represent the light curves binned in phase with a bin size of 0.002.
\label{fig:hatsouth}}
\ifthenelse{\boolean{emulateapj}}{
    \end{figure*}
}{
    \end{figure}
}
Operating since 2010, the HATSouth survey has discovered 16 exoplanets\footnote{For a complete list with reference of all the HATSouth planets, see \textit{http://hatsouth.org/}} so far, including the two presented in this paper. 
These planets span a range in masses that goes from super-Neptune \citep[HATS-7\,b and HATS-8\,b,][]{bakos:2015:hats7,bayliss:2015:hats8} to super-Jupiters \citep[e.g. HATS-11\,b and HATS-16\,b,][this work]{rabus:2015}.   

The paper is organized as follows: in \refsecl{obs} we describe the observations and data reduction that allowed us to discover and confirm the planetary nature of HATS-15b and HATS-16b. In \refsecl{analysis} we outline the diverse steps of our analysis that brought us to discard the false positive scenarios and determine the stellar and planetary parameters of the two systems. Finally in \refsecl{discussion}, we discuss and summarize our findings.  


\section{Observations}
\label{sec:obs}
\subsection{Photometric detection}
\label{sec:detection}

HATS-15 is a $V=14.77$\,mag star, located in the Capricornus constellation, while HATS-16 (aka GSC 7516-00867) is a $V= 13.83$\,mag star in the Sculptor constellation.  Both the stars have been identified as planetary host candidates based on roughly six months of continuous photometric observations.    
Specifically, HATS-15 was in the field-of-view (FOV) of two different HATS-fields (G581 and G582), and was therefore observed from September 2009 to September 2010 with all the units in the three sites. %
HATS-16 was observed from June 2013 to December 2013 with the HS-2, HS-4 and HS-6 units (located at LCO, HESS and SSO respectively). The target was simultaneously observed with two different cameras in each site, as it was lying in the overlapping region of the FOV of two cameras. \reftabl{photobs} shows the details of the photometric observations, displaying the total number of photometric measurements, the observing cadence and other information.
 
The light curves of the two stars were obtained through aperture photometry from the properly calibrated (bias and dark subtracted and flat fielded) science frames following \cite{bakos:2013:hatsouth} and \cite{penev:2013:hats1}. The light curves were detrended, using the TFA algorithm \citep[Trend Filtering Algorithm; see][]{kovacs:2005:TFA}, and then searched for a periodical signal by fitting them with a box-shaped transit model \citep[Box Least Square; ][]{kovacs:2002:BLS}. We found that the HATS-15 photometry shows a periodic dip every 1.75 days, while in the HATS-16 light curve we detected a transit signal with a periodicity of 2.69 days.
The phase-folded light curves of both the planets are shown in \reffig{hatsouth}, and the data are provided in \reftabl{phfu}.

\ifthenelse{\boolean{emulateapj}}{
    \begin{deluxetable*}{llrrrr}
}{
    \begin{deluxetable}{llrrrr}
}
\tablewidth{0pc}
\tabletypesize{\scriptsize}
\tablecaption{
    Summary of photometric observations
    \label{tab:photobs}
}
\tablehead{
    \multicolumn{1}{c}{Instrument/Field\tablenotemark{a}} &
    \multicolumn{1}{c}{Date(s)} &
    \multicolumn{1}{c}{\# Images} &
    \multicolumn{1}{c}{Cadence\tablenotemark{b}} &
    \multicolumn{1}{c}{Filter} &
    \multicolumn{1}{c}{Precision\tablenotemark{c}} \\
    \multicolumn{1}{c}{} &
    \multicolumn{1}{c}{} &
    \multicolumn{1}{c}{} &
    \multicolumn{1}{c}{(sec)} &
    \multicolumn{1}{c}{} &
    \multicolumn{1}{c}{(mmag)}
}
\startdata
\sidehead{\textbf{HATS-15}}
~~~~HS-1.3/G581 & 2009 Aug--2010 Sep & 6802~ & 288~~~ & $r$~~ & 19.1~~~~ \\
~~~~HS-3.3/G581 & 2009 Sep--2010 Sep & 8617~ & 292~~~ & $r$~~ & 17.9~~~~ \\
~~~~HS-5.3/G581 & 2009 Nov--2010 Sep & 586~ & 292~~~ & $r$~~ & 18.6~~~~ \\
~~~~HS-2.2/G582 & 2009 Sep--2010 Sep & 4450~ & 284~~~ & $r$~~ & 19.9~~~~ \\
~~~~HS-4.2/G582 & 2009 Sep--2010 Sep & 7834~ & 288~~~ & $r$~~ & 20.0~~~~ \\
~~~~HS-6.2/G582 & 2010 Aug--2010 Sep & 207~ & 290~~~ & $r$~~ & 21.7~~~~ \\
~~~~FTS~2\,m/Spectral & 2011 Sep 23     & 87~ & 82~~~ & $i$~~ & 2.1~~~~ \\
~~~~PEST~0.3\,m & 2013 May 21           & 143~ & 130~~~ & $R_{\mathrm{C}}$ & 12.5~~~~ \\
~~~~MPG~2.2\,m/GROND & 2013 Jun 15 & 71~ & 224~~~ & $g$~~ & 1.1~~~~ \\
~~~~MPG~2.2\,m/GROND & 2013 Jun 15 & 65~ & 224~~~ & $r$~~ & 0.8~~~~ \\
~~~~MPG~2.2\,m/GROND & 2013 Jun 15 & 70~ & 224~~~ & $i$~~ & 1.2~~~~ \\
~~~~MPG~2.2\,m/GROND & 2013 Jun 15 & 72~ & 224~~~ & $z$~~ & 1.5~~~~ \\
\sidehead{\textbf{HATS-16}}
~~~~HS-2.1/G588 & 2013 Jun--2013 Oct & 3888~ & 279~~~ & $r$~~ & 12.9~~~~ \\
~~~~HS-4.1/G588 & 2013 Jun--2013 Dec & 4683~ & 291~~~ & $r$~~ & 13.5~~~~ \\
~~~~HS-6.1/G588 & 2013 Jun--2013 Dec & 3618~ & 296~~~ & $r$~~ & 13.2~~~~ \\
~~~~HS-2.2/G588 & 2013 Jun--2013 Oct & 1929~ & 281~~~ & $r$~~ & 12.2~~~~ \\
~~~~HS-4.2/G588 & 2013 Jun--2013 Dec & 3732~ & 291~~~ & $r$~~ & 11.9~~~~ \\
~~~~HS-6.2/G588 & 2013 Jun--2013 Dec & 3678~ & 296~~~ & $r$~~ & 11.8~~~~ \\
~~~~DK~1.54\,m/DFOSC & 2014 Oct 06 & 170~ & 116~~~ & $R$~~ & 2.1~~~~ \\
\enddata
\tablenotetext{a}{
    For HATSouth data we list the HATSouth unit, CCD and field name
    from which the observations are taken. HS-1 and -2 are located at
    Las Campanas Observatory in Chile, HS-3 and -4 are located at the
    H.E.S.S. site in Namibia, and HS-5 and -6 are located at Siding
    Spring Observatory in Australia. Each unit has 4 ccds. Each field
    corresponds to one of 838 fixed pointings used to cover the full
    4$\pi$ celestial sphere. All data from a given HATSouth field and
    CCD number are reduced together, while detrending through External
    Parameter Decorrelation (EPD) is done independently for each
    unique unit+CCD+field combination.
}
\tablenotetext{b}{
    The median time between consecutive images rounded to the nearest
    second. Due to factors such as weather, the day--night cycle,
    guiding and focus corrections the cadence is only approximately
    uniform over short timescales.
}
\tablenotetext{c}{
    The RMS of the residuals from the best-fit model.
}
\ifthenelse{\boolean{emulateapj}}{
    \end{deluxetable*}
}{
    \end{deluxetable}
}

\ifthenelse{\boolean{emulateapj}}{
    \begin{deluxetable*}{llrrrrl}
}{
    \begin{deluxetable}{llrrrrl}
}
\tablewidth{0pc}
\tablecaption{
    Light curve data for HATS-15 and HATS-16\label{tab:phfu}.
}
\tablehead{
    \colhead{Object\tablenotemark{a}} &
    \colhead{BJD\tablenotemark{b}} & 
    \colhead{Mag\tablenotemark{c}} & 
    \colhead{\ensuremath{\sigma_{\rm Mag}}} &
    \colhead{Mag(orig)\tablenotemark{d}} & 
    \colhead{Filter} &
    \colhead{Instrument} \\
    \colhead{} &
    \colhead{\hbox{~~~~(2,400,000$+$)~~~~}} & 
    \colhead{} & 
    \colhead{} &
    \colhead{} & 
    \colhead{} &
    \colhead{}
}
\startdata
HATS-15 &  $ 55435.70436 $ & $  -0.02268 $ & $   0.01260 $ & $ \cdots $ & $ r$ &         HS\\
HATS-15 &  $ 55374.54242 $ & $  -0.00238 $ & $   0.02157 $ & $ \cdots $ & $ r$ &         HS\\
HATS-15 &  $ 55386.77495 $ & $   0.01740 $ & $   0.01000 $ & $ \cdots $ & $ r$ &         HS\\
HATS-15 &  $ 55096.69186 $ & $   0.02154 $ & $   0.01436 $ & $ \cdots $ & $ r$ &         HS\\
HATS-15 &  $ 55416.48228 $ & $   0.00679 $ & $   0.00988 $ & $ \cdots $ & $ r$ &         HS\\
HATS-15 &  $ 55124.65174 $ & $  -0.04501 $ & $   0.01408 $ & $ \cdots $ & $ r$ &         HS\\
HATS-15 &  $ 55110.67192 $ & $  -0.00283 $ & $   0.02212 $ & $ \cdots $ & $ r$ &         HS\\
HATS-15 &  $ 55367.55285 $ & $   0.02756 $ & $   0.00964 $ & $ \cdots $ & $ r$ &         HS\\
HATS-15 &  $ 55381.53294 $ & $   0.00608 $ & $   0.01109 $ & $ \cdots $ & $ r$ &         HS\\
HATS-15 &  $ 55409.49282 $ & $  -0.00161 $ & $   0.01282 $ & $ \cdots $ & $ r$ &         HS\
\enddata
\tablenotetext{a}{
    Either HATS-15, or HATS-16.
}
\tablenotetext{b}{
    Barycentric Julian Date is computed directly from the UTC time
    without correction for leap seconds.
}
\tablenotetext{c}{
The out-of-transit level has been subtracted. For observations made with the HATSouth instruments (identified by ``HS'' in the ``Instrument'' column) these magnitudes have been corrected for
trends using the EPD and TFA procedures applied {\em prior} to fitting the transit model. This procedure may lead to an artificial dilution in the transit depths. For HATS-15 the transit depth is 93\% and 100\% that of the true depth for the G581.3 and G582.2 observations, respectively. For HATS-16 it is 88\% and 87\% that of the true depth for the G588.1 and G588.2 observations, respectively. For observations made with follow-up instruments (anything other than ``HS'' in the ``Instrument'' column), the magnitudes have been corrected for a quadratic trend in time fit simultaneously with the transit. For HATS-16 an additional correction has been made for trends correlated with variations in the FWHM of the PSF.
}
\tablenotetext{d}{ Raw magnitude values without correction for the quadratic trend in time, or for trends correlated with the seeing. These are only reported for the follow-up observations.
}
\tablecomments{This table is available in a machine-readable form in the on-line journal. A portion is shown here for guidance regarding its form and content.}
\ifthenelse{\boolean{emulateapj}}{
    \end{deluxetable*}
}{
    \end{deluxetable}
}

\subsection{Spectroscopic Observations}
\label{sec:obsspec}

In order to confirm the planetary nature of the two candidates and obtain the complete set of their orbital and physical parameters, systematic spectroscopic observations of the two systems are mandatory. Our observations can be divided in two different steps: first we observed our targets with low resolution spectrographs, or at high resolution but low S/N, to obtain an initial characterization of the star and exclude some of the most probable false positive scenarios; subsequently, we obtained several high-resolution high S/N spectra to measure the radial velocity of the two stars.
\ifthenelse{\boolean{emulateapj}}{
    \begin{figure*} [!htbp]
}{
    \begin{figure}[!htbp]
}
\plottwo{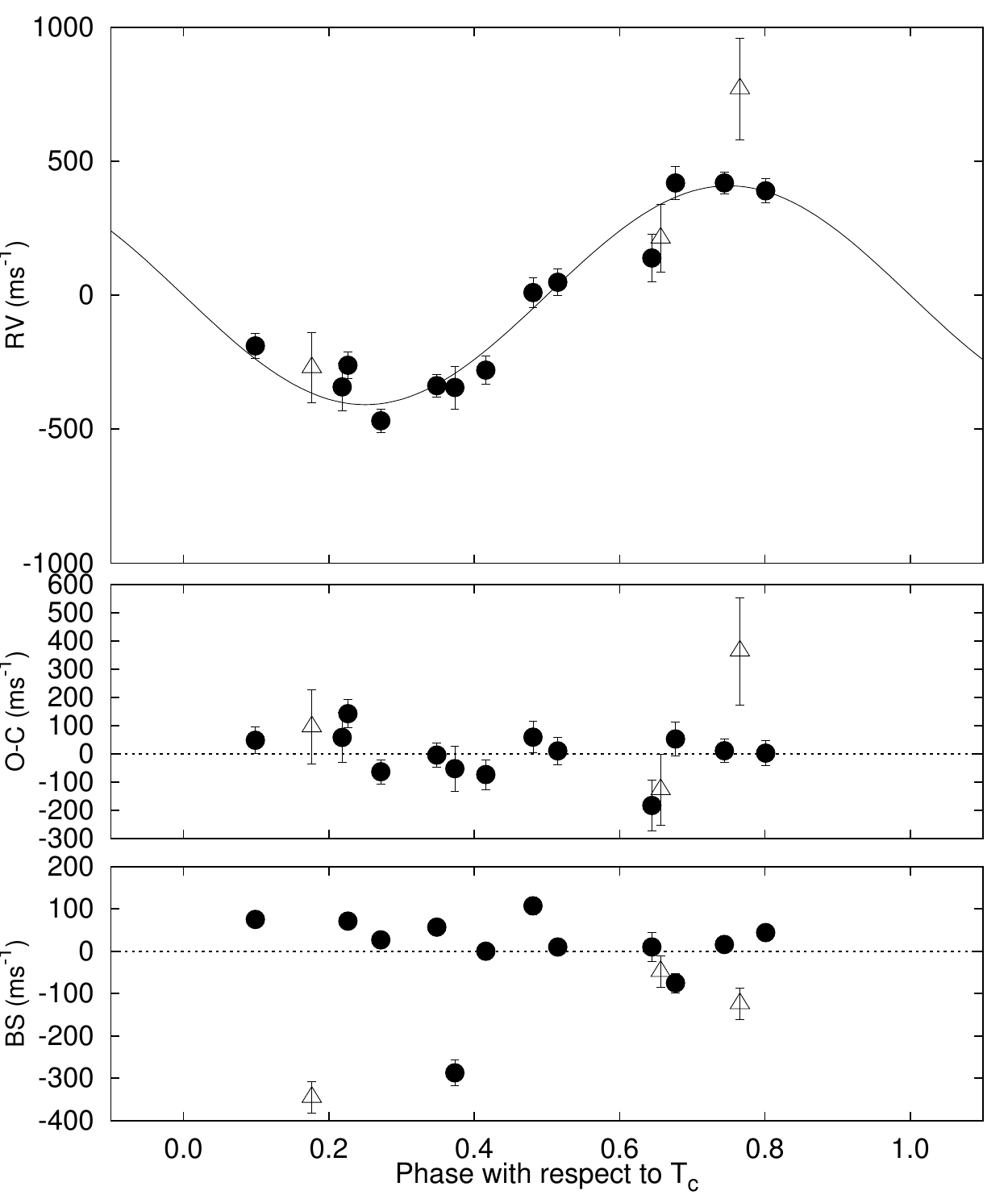}{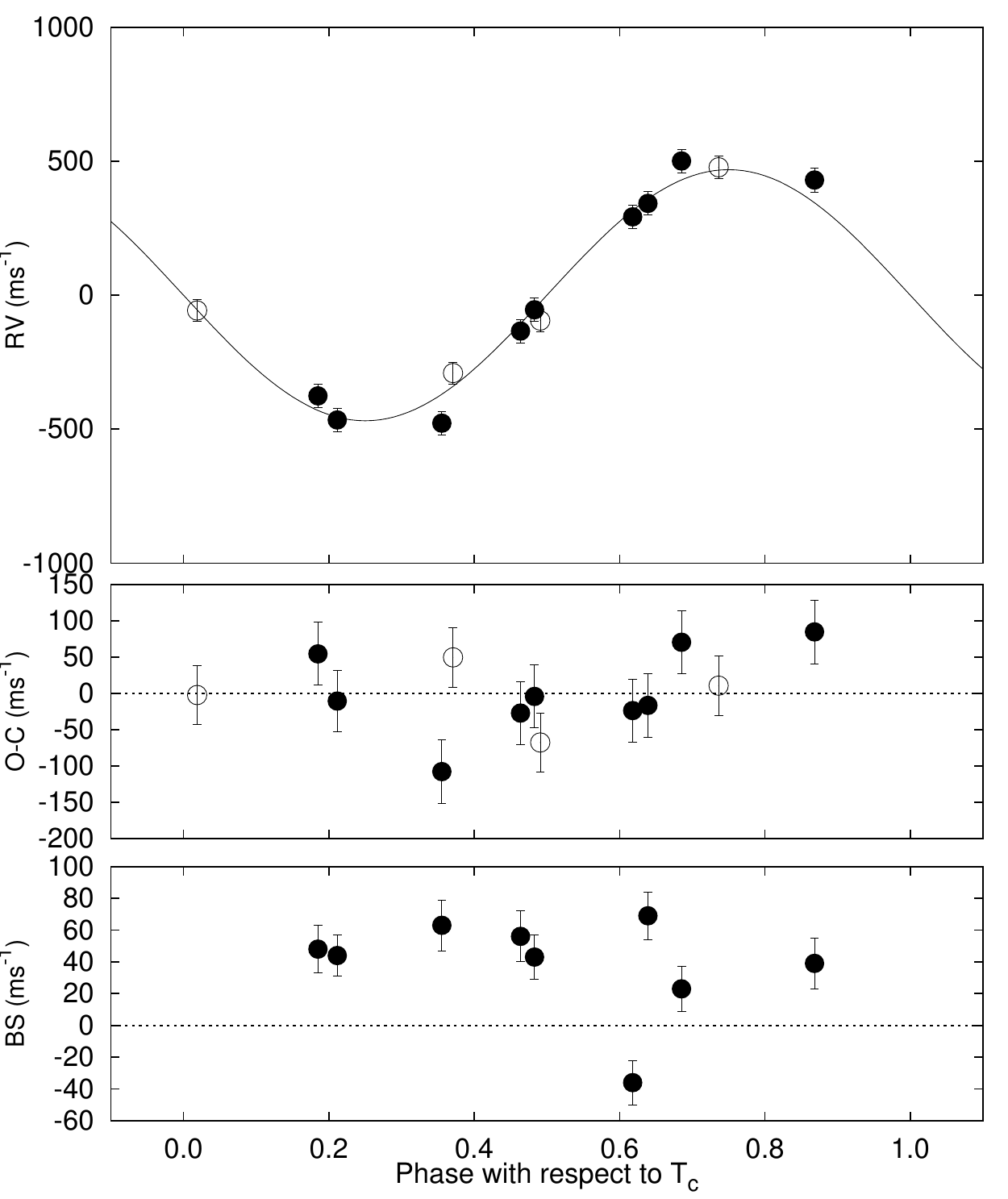}
\caption{
    Phased high-precision RV measurements for \hbox{HATS-15{}} (left), and \hbox{HATS-16{}} (right) from FEROS (filled circles), Coralie (open triangles), and HIRES (open circles). The top panel shows the phased measurements together with our best-fit circular-orbit model (see \reftabl{planetparam}) for each system. Zero-phase corresponds to the time of mid-transit and the center-of-mass velocity has been subtracted. The second panel shows the velocity $O\!-\!C$ residuals from the best fit. The error bars include the jitter terms listed in \reftabl{planetparam} added in quadrature to the formal errors for each instrument. The third panel shows the bisector spans (BS). Note the different vertical scales of the panels.
}
\label{fig:rvbis}
\ifthenelse{\boolean{emulateapj}}{
    \end{figure*}
}{
    \end{figure}
}

For both the stars we obtained 4 spectra with the Wide-Field Spectrograph (WiFeS) mounted on the 2.3\,m ANU telescope located at SSO. WiFeS is an image-slicing integral-field spectrograph \citep{dopita:2007}. According to the different slits used, it can achieve an average resolution up to $R=\lambda / \Delta \lambda \sim$7000 over a wavelength rage of 3300--9200\,\AA. The spectra were extracted and reduced following \cite{bayliss:2013:hats3}.
One of the spectra, was observed with a resolution of $R=3000$ for obtaining a first spectral classification of each star and verifying that they are not giants. By analyzing the spectra, we found that HATS-15 is a G9 dwarf ($\mathrm{T_{eff,*}}=5000 \pm 300$\,K and $\log{g_{\star}}=3.5 \pm 0.3$) while HATS-16 is a G3 dwarf ($\mathrm{T_{eff,*}}=6300 \pm 300$\,K and $\log{g_{\star}}= 4.6\pm 0.3$). 
The other three spectra where obtained with a higher resolution ($R=7000$) to look for a possible periodic RV signal higher than 5\,km\,s$^{-1}$, which is the signature of an eclipsing binary that is the most probable false positive scenario. In both the cases, we did not find evidence of any RV variation higher than 0.5\,km\,s$^{-1}$ for HATS-15 and 2\,km\,s$^{-1}$ for HATS-16. Moreover, none of the two systems show a composed spectrum, allowing us to exclude the spectroscopic binary scenario.
  
Another reconnaissance spectrum for HATS-15 was observed with the du~Pont telescope on August 21, 2013. The 2.5\,m du~Pont telescope is located at LCO, and is equipped with an echelle spectrograph capable to cover the optical range between 3700 and 7000\,\AA, achieving a maximum resolution of $R\sim$45000. We used a a $1^{\prime\prime} \times 4^{\prime\prime}$ slit that allowed a resolution of $R\sim$40000. The spectrum was reduced with an automated pipeline written for this instrument and similar to the one used for Coralie and FEROS \citep{brahm:2015:pipeline}.

The reconnaissance spectroscopy observations pointed towards a planetary system scenario for both the systems. Therefore, we started a campaign to get high precision RV measurements, in order to determine the mass of the companions, and verify that the periodical signals in RV were consistent with the photometric ones.

Between June and November 2012 we obtained three high resolution spectra of HATS-15 with Coralie. The spectrograph Coralie is fed from the 1.2\,m Swiss-Euler telescope, which is located at the ESO Observatory in La Silla, Chile. The instrument is a fiber-fed spectrograph with a resolution of $R\sim$60000 and a wavelength coverage between 3850 and 6900\,\AA \ \citep{queloz:2001}. During each science exposure, a simultaneous ThAr spectra was taken, to be able to operate a proper wavelength calibration. The spectra extraction was performed from the images calibrated with bias and flats obtained during the twilight following \cite{marsh:1989}. The RV of each spectrum was then measured by cross-correlation with a binary mask accurately chosen according to the spectral class of the target \citep[for a complete description of the reduction pipeline see][]{jordan:2014:hats4}.         

The bulk of the RV measurements, which allowed us to verify the planetary nature of both the candidates, was performed by utilizing the Fiber-fed Extended Range Optical Spectrograph (FEROS). It is an echelle spectrograph mounted on the MPG 2.2\,m telescope, which is also situated at the La Silla Observatory. The instrument is capable of a wide wavelength coverage from 3700 to 8600\,\AA\ and has an average resolution of $R\sim$48000 \citep{kaufer:1998}.  
For HATS-15 we observed 13 spectra between September 2011 and May 2013, whereas 12 spectra were obtained for HATS-16 between June and October 2014.
The data were reduced and RVs were measured, using the same pipeline written for Coralie and adapted for FEROS \citep{brahm:2015:pipeline}.

Moreover, we obtained four observations of HATS-16 with HIRES in September 2014. HIRES (HIgh Resolution Echelle Spectrometer) is an echelle spectrograph mounted in the Nasmyth focus of the Keck-I telescope located on the mountain of Mauna Kea in Hawaii, USA. HIRES has a resolution up to 84000 and can cover wavelengths between 3000 and 11000\,\AA\ \citep{vogt:1994}.
The observations were made using the standard high-precision RV setup for faint targets. We used the C2 decker obtaining a resolution of $R\sim 55000$.
Unlike the previous two instruments used for high-resolution spectroscopy, the wavelength calibration was not conducted with the simultaneous observation of a ThAr spectrum, but with an Iodine absorption cell \citep{marcy:1992}.
The radial velocity extraction was performed using a theoretical synthetic template drawn from the \citet{coelho:2014} grid as described by \citet{fulton:2015}.  This method provides greater precision measurements for faint stars than the traditional technique of obtaining an additional iodine-free observation to be used as a template \citep[e.g.,][]{bayliss:2015:hats8}.  

The high resolution RV measurements of both the systems showed a periodicity consistent with the one measured form the photometric observations.
A periodic signal, presented both in the light curve and in the RV, might not be necessarily due to the presence of a planetary companion, but can be produced by a blended eclipsing binary. To exclude this false positive scenario we measured the bisector span (BS) for both the systems finding no hint of a periodicity consistent with the orbital period; for the BS measurements we followed a procedure similar to that of \cite{torres:2007:hat3}, appropriately adapted to each instrument. 
The spectroscopic observations are summarized in \reftabl{specobs}, and phased high-precision RV and BS measurements are shown for each system in \reffigl{rvbis}. 
\ifthenelse{\boolean{emulateapj}}{
    \begin{figure*} [!htbp]
}{
    \begin{figure}[!htbp]
}
\plotone{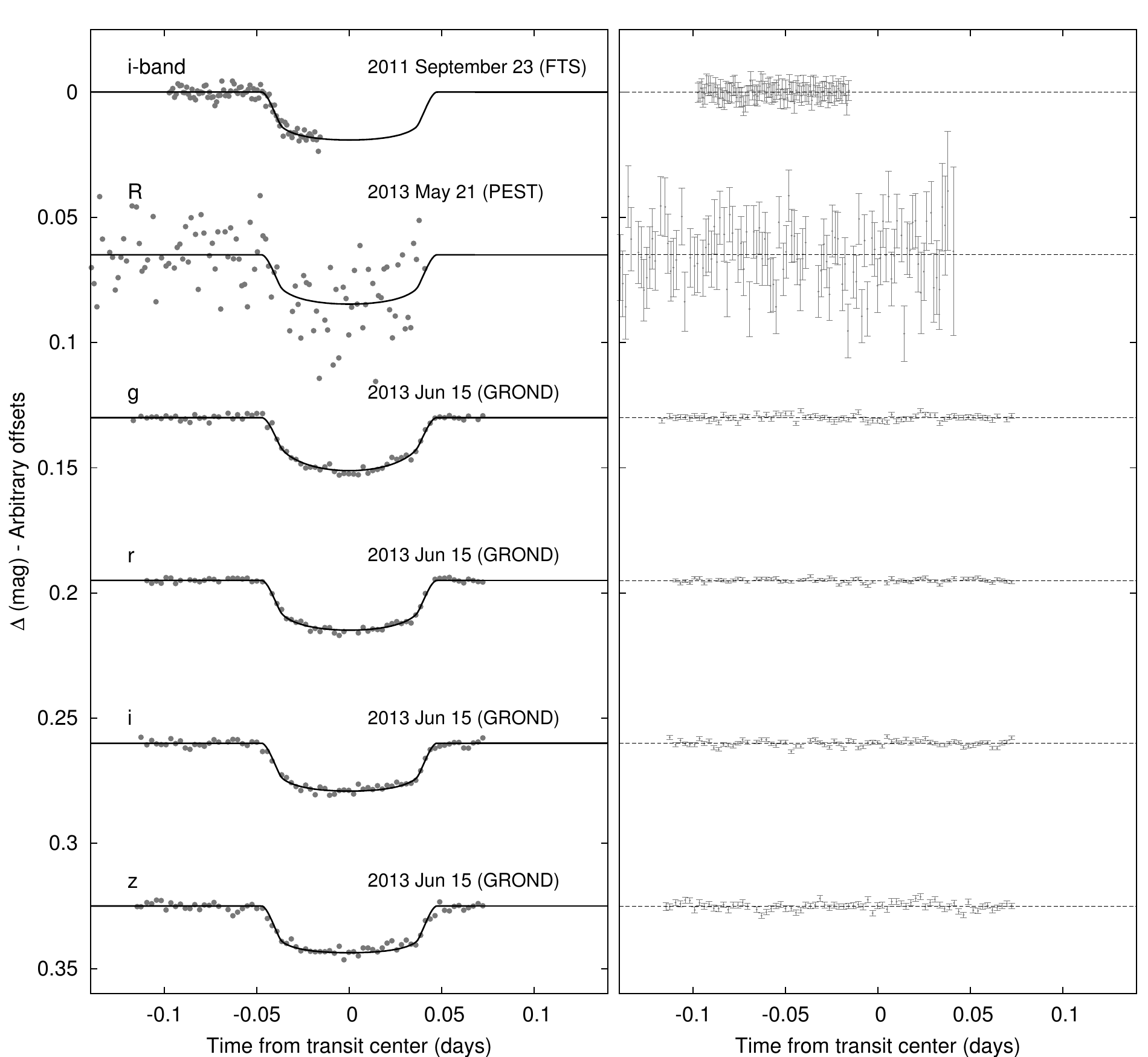}
\caption{
Left: Unbinned transit light curves for HATS-15.  The light curves have been corrected for quadratic trends in time fitted
simultaneously with the transit model.
The dates of the events, filters and instruments used are indicated.  Light curves following the first are displaced vertically for clarity.  Our best fit from the global modeling described in \refsecl{globmod} is shown by the solid lines.
Right: residuals from the fits are displayed in the same order as the left curves.  The error bars represent the photon and background shot noise, plus the readout noise.
}
\label{fig:lc15}
\ifthenelse{\boolean{emulateapj}}{
    \end{figure*}
}{
    \end{figure}
}
\setcounter{planetcounter}{2}
%
\ifthenelse{\boolean{emulateapj}}{
    \begin{figure} [!htbp]
}{
    \begin{figure}[!htbp]
}
\plotone{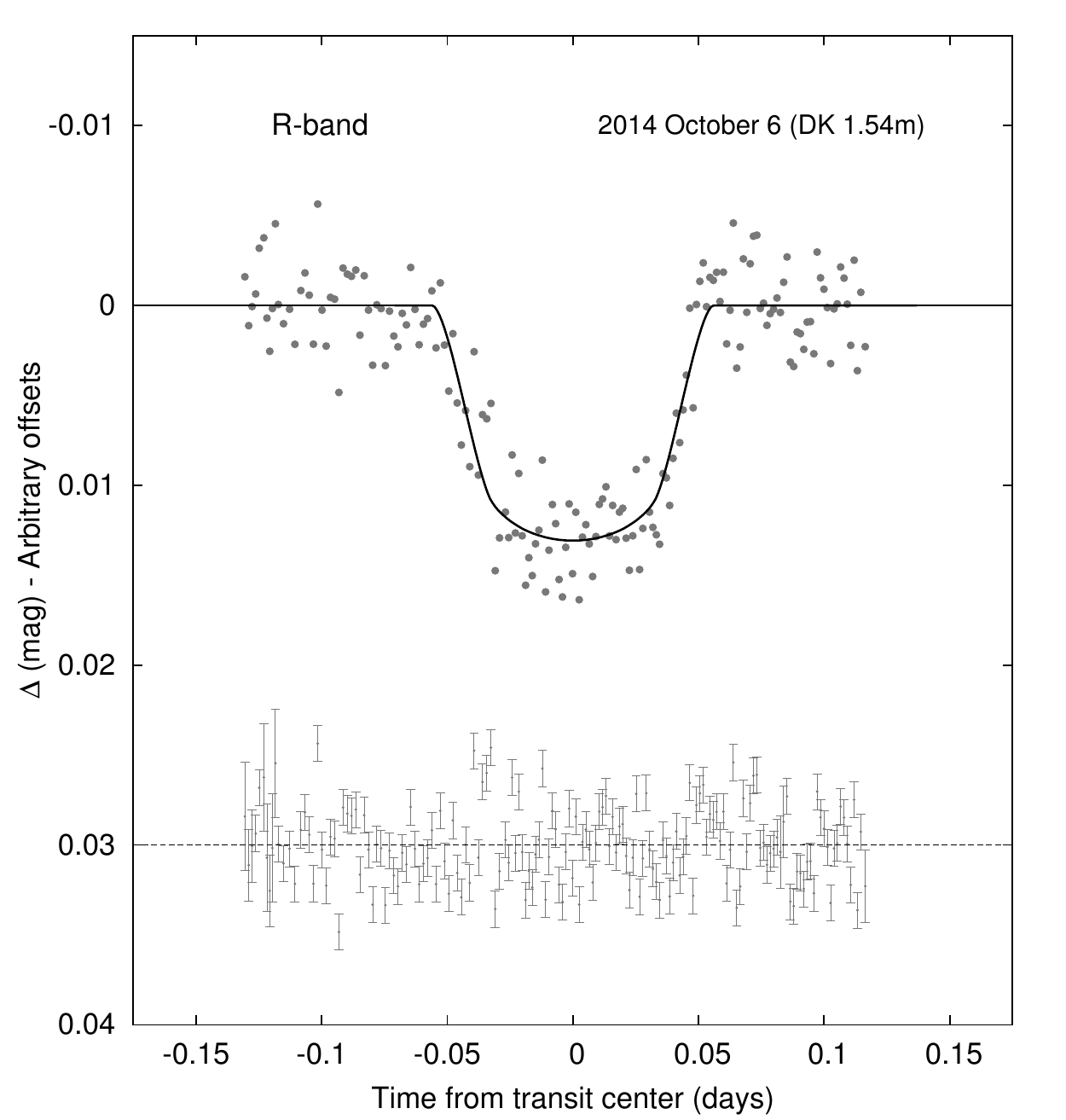}
\caption{
Similar to \reffigl{lc15}; we show the follow-up light curve for HATS-16. In this case, variations in the light curve that are correlated with the FWHM of the PSF are corrected simultaneously to the fitting of the transit.
}
\label{fig:lc16}
\ifthenelse{\boolean{emulateapj}}{
    \end{figure}
}{
    \end{figure}
}

Finally, four further high--resolution spectra of HATS-15 were obtained in August-September 2014 with the Planet Finder Spectrograph (PFS). PFS is an echelle spectrograph mounted on the 6.5\,m Magellan Clay Telescope at Las Campanas Observatory in Chile. The maximum wavelength coverage PSF is capable of span the optical region from 3880 to 6680\,\AA, and can be used in different modes, yielding a maximum resolution of $R\sim 190000$ \citep{crane:2010}. For our observations, we used a $0.5^{\prime\prime} \times 2.5^{\prime\prime}$ slit having a resolving power of 76000.
These spectra were used to determine the spectroscopic parameters for HATS-15 using the ZASPE program (see \refsecl{analysis}), while for HATS-16 the parameters have been measured by applying the same program to a combination of the FEROS spectra.  

\ifthenelse{\boolean{emulateapj}}{
    \begin{deluxetable*}{llrrrrr}
}{
    \begin{deluxetable}{llrrrrrrrr}
}
\tablewidth{0pc}
\tabletypesize{\scriptsize}
\tablecaption{
    Summary of spectroscopy observations
    \label{tab:specobs}
}
\tablehead{
    \multicolumn{1}{c}{Instrument}          &
    \multicolumn{1}{c}{UT Date(s)}             &
    \multicolumn{1}{c}{\# Spec.}   &
    \multicolumn{1}{c}{Res.}          &
    \multicolumn{1}{c}{S/N Range\tablenotemark{a}}           &
    \multicolumn{1}{c}{$\gamma_{\rm RV}$\tablenotemark{b}} &
    \multicolumn{1}{c}{RV Precision\tablenotemark{c}} \\
    &
    &
    &
    \multicolumn{1}{c}{$\Delta \lambda$/$\lambda$/1000} &
    &
    \multicolumn{1}{c}{($\mathrm{km\,s}^{-1}$)}              &
    \multicolumn{1}{c}{($\mathrm{m\,s}^{-1}$)}
}
\startdata
\sidehead{\textbf{HATS-15}}
\sidehead{Reconnaissance}%
ANU~2.3\,m/WiFeS & 2011 Jul 26--28 & 3 & 7 & 13--22 & -54.7 & 4000 \\
ANU~2.3\,m/WiFeS & 2011 Jul 27 & 1 & 3 & 95 & $\cdots$ & $\cdots$ \\
du~Pont~2.5\,m/Echelle & 2013 Aug 21 & 1 & 40 & 36 & -53.8 & 500 \\
\sidehead{High resolution radial velocity}%
MPG~2.2\,m/FEROS & 2011 Sep--2013 May & 13 & 48 & 16--49 & -54.145 & 81 \\
Euler~1.2\,m/Coralie & 2012 Jun--Nov & 3 & 60 & 12--13 & -54.270 & 246 \\
Magellan~6.5\,m/PFS & 2014 Sep & 4 & 76 & 40 & $\cdots$ & $\cdots$ \\
\sidehead{\textbf{HATS-16}}
\sidehead{Reconnaissance}%
ANU~2.3\,m/WiFeS & 2014 Jun 3--5 & 3 & 7 & 21--82 & 28.7 & 4000 \\
ANU~2.3\,m/WiFeS & 2014 Jun 4 & 1 & 3 & 101 & $\cdots$ & $\cdots$ \\
Euler~1.2\,m/Coralie & 2014 Jun 20 & 1 & 60 & 14 & 31.625 & 100 \\
\sidehead{High resolution radial velocity}%
MPG~2.2\,m/FEROS & 2014 Jun--Oct & 12 & 48 & 35--57 & 31.594 & 54 \\
Keck~10\,m/HIRES & 2014 Jun 20 & 4 & 55 & $\cdots$ & $\cdots$ & 20 \\
\enddata 
\tablenotetext{a}{
    S/N per resolution element near 5180\,\AA.
}
\tablenotetext{b}{
For the Coralie and FEROS observations of HATS-15, and for the FEROS observations of HATS-16, this is the zero-point RV from the best-fit orbit. For the WiFeS and du~Pont Echelle, and for the single Coralie observation of HATS-16 it is the mean value. We do not provide this quantity for the lower resolution WiFeS observations which were only used to measure stellar atmospheric parameters, or for the PFS observations of HATS-15 which were obtained without the I$_{2}$ cell and were used to determine the atmospheric parameters of the star.
}
\tablenotetext{c}{
For High-precision RV observations included in the orbit determination this is the scatter in the RV residuals from the best-fit orbit (which may include astrophysical jitter), for other instruments this is either an estimate of the precision (not including jitter), or the measured standard deviation. We do not provide this quantity for low-resolution observations from the ANU~2.3\,m/WiFeS, or for the PFS observations of HATS-15.
}
\ifthenelse{\boolean{emulateapj}}{
    \end{deluxetable*}
}{
    \end{deluxetable}
}

%
\setcounter{planetcounter}{1}
%

\subsection{Photometric follow-up observations}
\label{sec:phot}

Both the HATS-15 and HATS-16 planetary systems have been photometrically followed-up to properly constrain their orbital ephemeris and obtain a precise measure of the photometric parameters, which allow to directly estimate the mean density of the parent stars and, at the end, the radii of the planets. The photometric follow-up observations are summarized in \reftabl{photobs} and the corresponding data are reported in \reftabl{phfu} for both the systems.

We observed two partial transits and a complete one of HATS-15.
Specifically, the first half transit was obtained with the Faulkes Telescope South (FTS) on September the $23^{\mathrm{th}}$ 2011, and consisted in the out-of-transit data, ingress and partial transit observation. The FTS is a 2.0\,m telescope, located at SSO, and is part of the Las Cumbres Observatory Global Telescope (LCOGT) network. The telescope is equipped with a $4000 \times 4000$ pixels camera with a pixel size of 0.15$^{\prime\prime}$. The observations were performed with an \textit{i}-band filter and with the telescope out of focus, as no close-in background star was detected. The data reduction was performed using an automated aperture photometry pipeline
utilising Sextractor \citep{bertin:1996}. In brief the science images are first calibrated via bias subtraction and flat-fielding, then the light curves are
extracted via fix-aperture photometry.    

The second-half transit was observed with the 30\,cm PEST telescope on May the $21^{\mathrm{st}}$ 2013, and covered almost the whole transit excluding egress. The SBIG ST-8XME camera mounted on this telescope has a FOV of $31^{\prime} \times 21^{\prime}$, with a resolution of 1.2 arcsec per pixel. A $R_{\mathrm{C}}$ filter was used to take in-focus images with a cadence of 130\,s. The light curve was extracted from the calibrated images (dark subtracted and flat-fielded) with aperture photometry. For a more exhaustive overview on the instrument's characteristics and the data reduction description see \cite{zhou:2014:mebs}. 

A complete transit of HATS-15b was observed simultaneously in four different optical bands ($g^{\prime}$, $r^{\prime}$, $i^{\prime}$, $z^{\prime}$, similar to Sloan filters) using the multiband imager instrument GROND \citep[Gamma Ray Optical Near-infrared Detector;][]{greiner:2008}. The observations, performed on June the $15^{\mathrm{th}}$ 2013, were obtained with the telescope slightly out of focus to increase the photometric precision. After de-biasing and flat-fielding the science frames, the photometry was extracted with a pipeline based on DAOPHOT that make use of the APER IDL function \citep[for the GROND observing strategy and subsequent data reduction refer to][]{penev:2013:hats1, mohlerfischer:2013:hats2}. 
The light curves for all the HATS-15 follow-up transits are shown in \reffigl{lc15}. 

We obtained a transit of HATS-16 with the DFOSC camera on October the $6^{\mathrm{th}}$ 2014. The DFOSC (Danish Faint Object Spectrograph and Camera) instrument is mounted  on the 1.54\,m Danish Telescope, at the La Silla Observatory. The CCD has a resolution of 0.39 arcsec per pixel and a FOV $13.7^{\prime} \times 13.7^{\prime}$.  The telescope was defocused and a Bessel $R$ filter was used. After properly calibrating the images, aperture photometry was done following \cite{deeg:2001} to obtain the light curve. See \cite{rabus:2015} for a complete description of the instrument and reduction pipeline.
The light curve for HATS-16 is shown in \reffigl{lc16}.

\setcounter{planetcounter}{1}
\ifthenelse{\boolean{emulateapj}}{
    \begin{deluxetable*}{lccl}
}{
    \begin{deluxetable}{lccl}
}
\tablewidth{0pc}
\tabletypesize{\scriptsize}
\tablecaption{
    Stellar parameters for HATS-15 and HATS-16
    \label{tab:stellar}
}
\tablehead{
    \multicolumn{1}{c}{} &
    \multicolumn{1}{c}{\bf HATS-15} &
    \multicolumn{1}{c}{\bf HATS-16} &
    \multicolumn{1}{c}{} \\
    \multicolumn{1}{c}{~~~~~~~~Parameter~~~~~~~~} &
    \multicolumn{1}{c}{Value}                     &
    \multicolumn{1}{c}{Value}                     &
    \multicolumn{1}{c}{Source}
}
\startdata
\noalign{\vskip -3pt}
\sidehead{Astrometric properties and cross-identifications}
~~~~2MASS-ID\dotfill                & 2MASS 20442207-1926150  & 2MASS 23541409-3000467 & \\
~~~~GSC-ID\dotfill                  & ---      & GSC 7516-00867     & \\
~~~~R.A. (J2000)\dotfill            & \ensuremath{20^{\mathrm h}44^{\mathrm m}22.20^{\mathrm s}}      & \ensuremath{23^{\mathrm h}54^{\mathrm m}14.04^{\mathrm s}}    & 2MASS\\
~~~~Dec. (J2000)\dotfill            & \ensuremath{-19{\arcdeg}26{\arcmin}15.0{\arcsec}}      & \ensuremath{-30{\arcdeg}00{\arcmin}46.8{\arcsec}}   & 2MASS\\
~~~~$\mu_{\rm R.A.}$ (\masy)        & \ensuremath{19.6\pm2.5}    & \ensuremath{21.7\pm1.3} & UCAC4\\
~~~~$\mu_{\rm Dec.}$ (\masy)        & \ensuremath{3.7 \pm2.4}    & \ensuremath{-2.6\pm1.2} & UCAC4\\
\sidehead{Spectroscopic properties}
~~~~$\teffstar$ (K)\dotfill         &  \ensuremath{5311 \pm77   }   & \ensuremath{5738  \pm79   } & ZASPE\tablenotemark{a}\\
~~~~$\feh$\dotfill                  &  \ensuremath{0.000\pm0.050}   & \ensuremath{−0.100\pm0.050} & ZASPE               \\
~~~~$\vsini$ ($\mathrm{km\,s}^{-1}$)\dotfill         &  \ensuremath{4.18 \pm0.50 }   & \ensuremath{6.17  \pm0.22 } & ZASPE                \\
~~~~$\vmac$ ($\mathrm{km\,s}^{-1}$)\dotfill          &   3.3                 & 4.0                & Assumed\tablenotemark{b}              \\
~~~~$\vmic$ ($\mathrm{m\,s}^{-1}$)\dotfill           &   0.85                & 1.06               & Assumed\tablenotemark{c}              \\
~~~~$\gamma_{\rm RV}$ ($\mathrm{km\,s}^{-1}$)\dotfill&  \ensuremath{−54.145\pm0.020}  & \ensuremath{31.594\pm0.023} & FEROS\tablenotemark{d}  \\
\sidehead{Photometric properties}
~~~~$B$ (mag)\dotfill               &  \ensuremath{15.797\pm0.030}  & \ensuremath{14.477\pm0.060}  & APASS\tablenotemark{e} \\
~~~~$V$ (mag)\dotfill               &  \ensuremath{14.774\pm0.010}  & \ensuremath{13.834\pm0.020}  & APASS\tablenotemark{e} \\
~~~~$g$ (mag)\dotfill               &  \ensuremath{15.323\pm0.010}  & \ensuremath{14.086\pm0.020}  & APASS\tablenotemark{e} \\
~~~~$r$ (mag)\dotfill               &  \ensuremath{14.592\pm0.010}  & \ensuremath{13.645\pm0.010}  & APASS\tablenotemark{e} \\
~~~~$i$ (mag)\dotfill               &  \ensuremath{14.320\pm0.010}  & \ensuremath{13.496\pm0.030}  & APASS\tablenotemark{e} \\
~~~~$J$ (mag)\dotfill               &  \ensuremath{13.261\pm0.027}  & \ensuremath{12.652\pm0.024}  & 2MASS           \\
~~~~$H$ (mag)\dotfill               &  \ensuremath{12.806\pm0.024}  & \ensuremath{12.335\pm0.025}  & 2MASS           \\
~~~~$K_s$ (mag)\dotfill             &  \ensuremath{12.724\pm0.032}  & \ensuremath{12.280\pm0.021}  & 2MASS           \\
\sidehead{Derived properties}
~~~~$\mstar$ ($\msun$)\dotfill      &  \ensuremath{0.871\pm0.023}  &  \ensuremath{0.970\pm0.035}  & YY+$\rhostar$+ZASPE \tablenotemark{f}\\
~~~~$\rstar$ ($\rsun$)\dotfill      &  \ensuremath{0.922\pm0.027}  &  \ensuremath{1.238^{+0.097}_{−0.127}}  & YY+$\rhostar$+ZASPE         \\
~~~~$\loggstar$ (cgs)\dotfill       &  \ensuremath{4.449\pm0.022}  &  \ensuremath{4.239\pm0.079}  & YY+$\rhostar$+ZASPE         \\
~~~~$\rhostar$ (\gcmc)\dotfill      &  \ensuremath{1.57 \pm0.12 }  &  \ensuremath{0.72^{+0.26}_{−0.13}}  & YY+$\rhostar$+ZASPE \tablenotemark{g}         \\
~~~~$\lstar$ ($\lsun$)\dotfill      &  \ensuremath{0.625\pm0.057}  &  \ensuremath{\pm}  & YY+$\rhostar$+ZASPE         \\
~~~~$M_V$ (mag)\dotfill             &  \ensuremath{5.43 \pm0.11 }  &  \ensuremath{\pm}  & YY+$\rhostar$+ZASPE         \\
~~~~$M_K$ (mag,ESO)\dotfill &  \ensuremath{3.556\pm0.071}  &  \ensuremath{\pm}  & YY+$\rhostar$+ZASPE         \\
~~~~Age (Gyr)\dotfill               & \ensuremath{11.0^{+1.4}_{−2.0}}  &  \ensuremath{9.5  \pm1.8  }  & YY+$\rhostar$+ZASPE         \\
~~~~$A_{V}$ (mag)\dotfill           &  \ensuremath{0.151\pm0.063}  &  \ensuremath{0.000\pm0.021}  & YY+$\rhostar$+ZASPE         \\
~~~~Distance (pc)\dotfill           &  \ensuremath{689  \pm23   }  &  \ensuremath{774  \pm74   }  & YY+$\rhostar$+ZASPE\\ 
~~~~$P_{\mathrm{rot \star}}$ (d)\dotfill  &  $\cdots$              &  $12.350\pm0.024 $   & HATS light curves\\ [-1.5ex]\\
\enddata
\tablecomments{
For HATS-15\,b the fixed-circular-orbit
model has a Bayesian evidence that is $\sim 2$ times larger than the
evidence for the eccentric-orbit model, while for HATS-16\,b the fixed-circular-orbit model has a Bayesian evidence that is $\sim 6$ times larger than the evidence for the eccentric-orbit model. We therefore assume a fixed circular orbit in generating the parameters listed here.
}
\tablenotetext{a}{
    ZASPE = Zonal Atmospherical Stellar Parameter Estimator routine
    for the analysis of high-resolution spectra \citep{brahm:2015:pipeline},
    applied to the PFS and FEROS spectra of HATS-15 and HATS-16 respectively. These
    parameters rely primarily on ZASPE, but have a small dependence
    also on the iterative analysis incorporating the isochrone search
    and global modeling of the data.
}
\tablenotetext{b}{ The macro-turbulence values are obtained using the relations presented in \cite{valenti:2005}
}
\tablenotetext{c}{ The micro-turbulence values are computed interpolating the results reported in the SWEET-Cat catalogue \citep{santos:2013} for the relative T$_{\mathrm{eff}}$ and log\,\textit{g}.
}
\tablenotetext{d}{
    The error on $\gamma_{\rm RV}$ is determined from the orbital fit to
    the FEROS RV measurements, and does not include the systematic
    uncertainty in transforming the velocities from FEROS to the IAU
    standard system. The velocities have not been corrected for gravitational redshifts.
} \tablenotetext{e}{
    From APASS DR6 for HATS-15, HATS-16 as
    listed in the UCAC 4 catalog \citep{zacharias:2012:ucac4}.  
}
\tablenotetext{f}{
    YY+\rhostar+ZASPE = Based on the YY isochrones \citep{yi:2001}, \rhostar\ as a luminosity indicator, and the ZASPE results.
}
\tablenotetext{g}{
    In the case of $\rhostar$ the parameter is primarily determined
    from the global fit to the light curves and RV data. The value
    shown here also has a slight dependence on the stellar models and
    ZASPE parameters due to restricting the posterior distribution to
    combinations of $\rhostar$+$\teffstar$+$\feh$ that match to a
    YY stellar model.
}
\ifthenelse{\boolean{emulateapj}}{
    \end{deluxetable*}
}{
    \end{deluxetable}
}

\section{Analysis}
\label{sec:analysis}

\subsection{Properties of the parent star}
\label{sec:stelparam}

For achieving a precise determination of the physical properties of a new exoplanet, it is crucial to properly characterize its host star, especially its mass and radius. We obtained those quantities, and the other set of parameters describing the HATS-15 and HATS-16 stars, by properly combining the spectroscopic and photometric data.

The atmospheric parameters, including the effective temperature $T_{\mathrm{eff}}$, surface gravity $\log{g}$, metallicity [Fe/H] and the projected rotational velocity $v\sin{i}$, were measured by analyzing the high resolution spectra obtained with PFS for HATS-15 and FEROS for HATS-16. The analysis was performed using the Zonal Atmospherical  Stellar  Parameter  Estimator (ZASPE) code \citep[see][]{brahm:2015:pipeline}.
In brief, the atmospheric parameters are calculated iteratively, selecting a specific region of the spectra (between 5000\,\AA\,and 6000\,\AA) and fitting the median-combined observed spectra with a grid of synthetic ones \citep{husser:2013}. For HATS-15 we found: $T_{\mathrm{eff}}=5296\pm76$, $\log{g}=4.60\pm0.12$, [Fe/H]=$0.090\pm0.040$  and $v\sin{i}= 4.36\pm0.24$; while for HATS-16: $T_{\mathrm{eff}}=5840\pm120$, $\log{g}=4.50\pm0.19$, [Fe/H]=$ −0.010\pm0.070 $ and $v\sin{i}=6.01\pm0.50$.

The fundamental stellar parameters were obtained combining the spectroscopic and photometric quantities with the Yonsei-Yale stellar evolutionary models \citep[Y2 hereafter;][]{yi:2001}.
In particular, for the analysis with the isochrones, we used the stellar density $\rho_{\star}$ obtained from the photometry instead of $\log{g}$ from the spectra, as it provides a more precise and more accurate constraint on the stellar properties \citep[following][]{sozzetti:2007}. Assuming a nil eccentricity of the planetary orbit, the stellar density can be directly measured from the transit light curve as described in \cite{seager:2003}.

For both the systems we performed a second time the analysis: first we re-derived the spectroscopic quantities with ZASPE, this time fixing the values of the log\,\textit{g} to the ones retrieved from the stellar evolution models. Then we obtained the stellar properties once more modeling the data with the Y2 isochrones. We found that HATS-15 has a mass $M_{\star}=0.871\pm0.023\,M_{\sun}$, radius $R_{\star}=0.922\pm0.027\,R_{\sun}$ and an age of $11.0^{+1.4}_{-2.0}$\,Gyr; while HATS-16 has $M_{\star}=0.970\pm0.035\,M_{\sun}$, $R_{\star}=1.238^{+0.097}_{-0.127}\,R_{\sun}$ and age=$9.5\pm1.8$\,Gyr. 
The final values adopted in the subsequent analysis to derive the planetary parameters are presented in \reftabl{stellar}, while in \reffigl{iso} the two stars are shown in a $\teffstar$--$\rhostar$ diagram (similar to a Hertzsprung-Russell diagram).    

To measure the distance of the two planetary systems we compared the magnitude observed with each filter with a set of predicted ones. The predicted magnitudes were determined using the Y2 isochrones and assuming an extinction law with $R_{\mathrm{V}} = 3.1$ from \citet{cardelli:1989}.
Using the NASA/IPAC Extragalactic Database (NED\footnote{The NASA/IPAC Extragalactic Database (NED) is operated by the Jet Propulsion Laboratory, California Institute of Technology, under contract with the National Aeronautics and Space Administration.}) we checked that the value for the extinction were consistent with the expected reddening at the Galactic position of both the systems. We found that, within the error bars the $A_{\mathrm{V}}$ are consistent with those predicted from the extinction map by \cite{Schlegel:1998} and \cite{schlafly:2011}. HATS-15 is $690 \pm 23$\,pc distant from the Sun, and HATS-16 is slightly farther away at $774 \pm 74$\,pc.

\ifthenelse{\boolean{emulateapj}}{
    \begin{figure*}[!htbp]
}{
    \begin{figure}[!htbp]
}
\plottwo{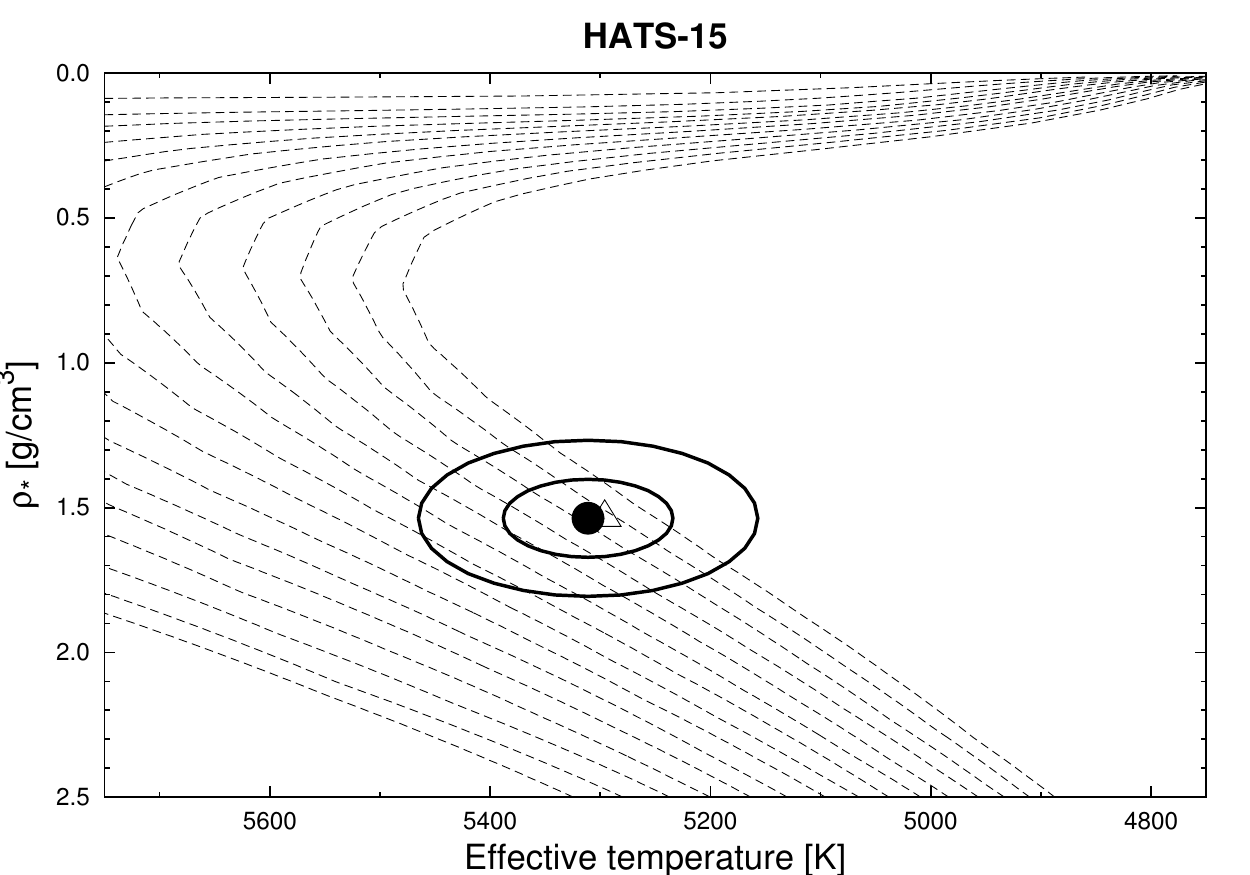}{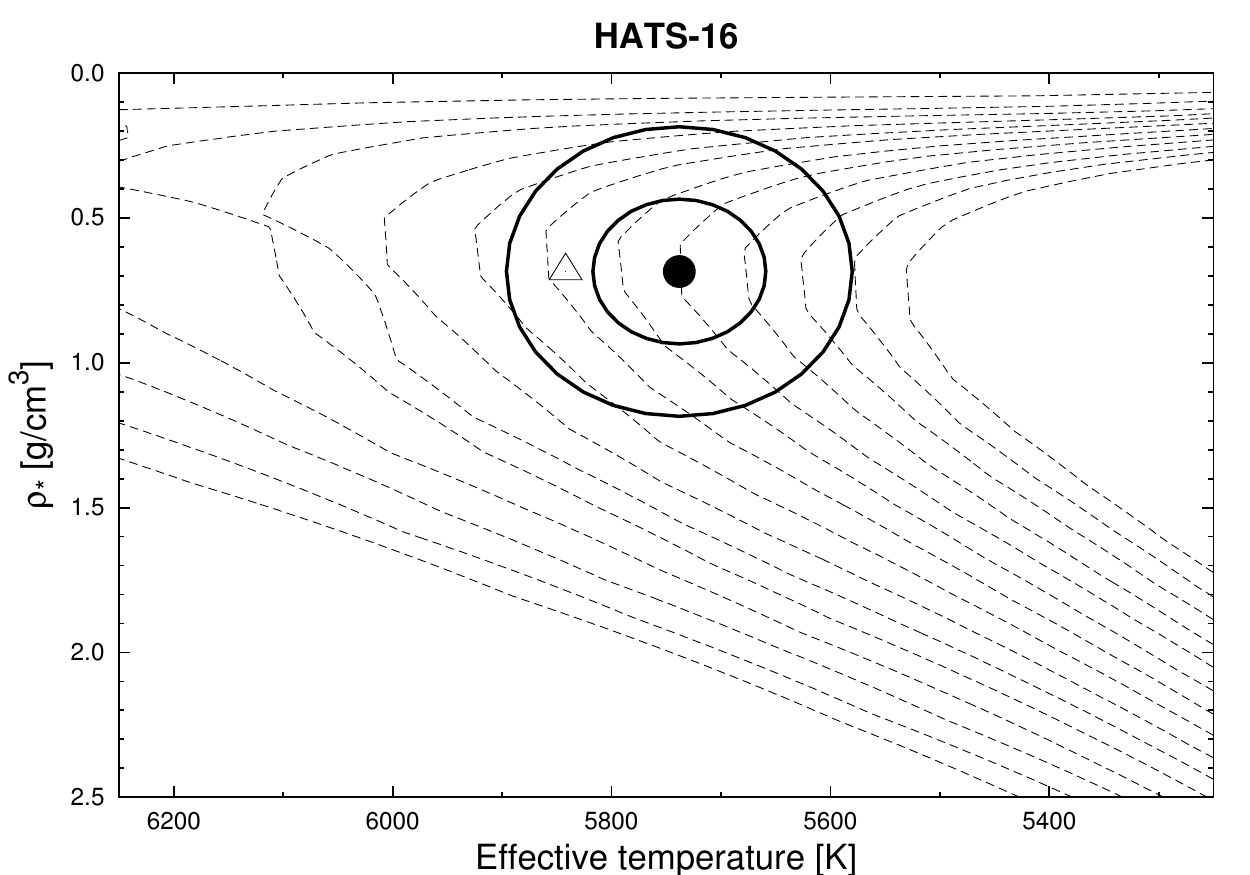}
\caption{Model isochrones from \cite{yi:2001} for the measured metallicities of HATS-15 (left) and HATS-16 (right). We show models for ages of 0.2\,Gyr and 1.0 to 14.0\,Gyr in 1.0\,Gyr increments (ages increasing from left to right). The adopted values of $\teffstar$ and \rhostar\ are shown together with their 1$\sigma$ and 2$\sigma$ confidence ellipsoids. The initial values of \teffstar\ and \rhostar\ from the first ZASPE and light curve analysis are represented with a triangle.}
\label{fig:iso}
\ifthenelse{\boolean{emulateapj}}{
    \end{figure*}
}{
    \end{figure}
}


%
\ifthenelse{\boolean{emulateapj}}{
    \begin{deluxetable*}{lcc}
}{
    \begin{deluxetable}{lcc}
}
\tabletypesize{\scriptsize}
\tablecaption{Orbital and planetary parameters for HATS-15\,b and HATS-16\,b\label{tab:planetparam}}
\tablehead{
    \multicolumn{1}{c}{} &
    \multicolumn{1}{c}{\bf HATS-15\,b} &
    \multicolumn{1}{c}{\bf HATS-16\,b} \\ 
    \multicolumn{1}{c}{~~~~~~~~~~~~~~~Parameter~~~~~~~~~~~~~~~} &
    \multicolumn{1}{c}{Value} &
    \multicolumn{1}{c}{Value}
}
\startdata
\noalign{\vskip -3pt}
\sidehead{Light curve parameters}
~~~$P$ (days)             \dotfill    &  \ensuremath{1.74748753\pm0.00000094}  &  \ensuremath{2.686502\pm0.000011}  \\
~~~$T_c$ (${\rm BJD}$)    
      \tablenotemark{a}   \dotfill    &  \ensuremath{2456387.21161\pm0.00015}  &  \ensuremath{2456824.79069\pm0.00076}   \\
~~~$T_{12}$ (days)
      \tablenotemark{a}   \dotfill    &  \ensuremath{0.09457\pm0.00074}  &  \ensuremath{0.1092v0.0032}   \\
~~~$T_{12} = T_{34}$ (days)
      \tablenotemark{a}   \dotfill    &  \ensuremath{0.01144\pm0.00069}  &  \ensuremath{0.0211\pm0.0044}   \\
~~~$\arstar$              \dotfill    &  \ensuremath{6.33\pm0.16}  &  \ensuremath{6.50^{+0.70}_{−0.43}}   \\
~~~$\zrstar$ \tablenotemark{b} \dotfill & \ensuremath{23.991^{+0.118}_{−0.081}} &  \ensuremath{22.27\pm0.49}   \\
~~~$\rpl/\rstar$          \dotfill    &  \ensuremath{0.1229\pm0.0012}  &  \ensuremath{0.1075\pm0.0037}   \\
~~~$b^2$                  \dotfill    & \ensuremath{0.100^{+0.044}_{−0.054}} & \ensuremath{0.536^{+0.065}_{−0.125}} \\
~~~$b \equiv a \cos i/\rstar$
                          \dotfill    & \ensuremath{0.317^{+0.064}_{−0.101}} & \ensuremath{0.732^{+0.043}_{−0.091}} \\
~~~$i$ (deg)              \dotfill    & \ensuremath{87.13^{+0.97}_{−0.67}} & \ensuremath{83.53^{+1.37}_{−0.86}}\phn \\

\sidehead{Limb-darkening coefficients \tablenotemark{c}}
~~~$c_1,g$ (linear term)    \dotfill    & 0.6795 & $\cdots$ \\
~~~$c_2,g$ (quadratic term) \dotfill    & 0.1376 & $\cdots$ \\
~~~$c_1,r$                  \dotfill    & 0.4560 & 0.3526\\
~~~$c_2,r$                  \dotfill    & 0.2619 & 0.3264\\
~~~$c_1,i$                  \dotfill    & 0.3470 & $\cdots$ \\
~~~$c_2,i$                  \dotfill    & 0.2834 & $\cdots$ \\
~~~$c_1,z$                  \dotfill    & 0.2742 & $\cdots$ \\
~~~$c_2,z$                  \dotfill    & 0.2937 & $\cdots$ \\
~~~$c_1,R$                  \dotfill    & 0.4259 & 0.3289\\
~~~$c_2,R$                  \dotfill    & 0.2687 & 0.3276\\
\sidehead{RV parameters}
~~~$K$ ($\mathrm{m\,s}^{-1}$)              \dotfill    &   \ensuremath{399 \pm26}  &  \ensuremath{485 \pm25}\phn\phn \\
~~~$e$ \tablenotemark{d}               \dotfill    & \ensuremath{< 0.126} & \ensuremath{< 0.000} \\

\sidehead{Planetary parameters}
~~~$\mpl$ ($\mjup$)       \dotfill    &   \ensuremath{2.17 \pm0.15 }  &  \ensuremath{3.27\pm0.19} \\
~~~$\rpl$ ($\rjup$)       \dotfill    &   \ensuremath{1.105\pm0.040}  &  \ensuremath{1.30\pm0.15} \\
~~~$C(\mpl,\rpl)$
    \tablenotemark{e}     \dotfill    &   \ensuremath{0.18}  &  \ensuremath{0.10} \\
~~~$\rhopl$ (\gcmc)       \dotfill    &   \ensuremath{1.97   \pm0.24   }  &  \ensuremath{1.86^{+0.94}_{−0.48}} \\
~~~$\log g_p$ (cgs)       \dotfill    &   \ensuremath{3.641  \pm0.040  }  &  \ensuremath{3.685^{+0.117}_{−0.086}} \\
~~~$a$ (AU)               \dotfill    &   \ensuremath{0.02712\pm0.00023}  &  \ensuremath{0.03744\pm0.00045} \\
~~~$T_{\rm eq}$ (K)        \dotfill   &   \ensuremath{1505   \pm30     }  &  \ensuremath{1592^{+61}_{−82}} \\
~~~$\Theta$ \tablenotemark{f} \dotfill &  \ensuremath{0.1211 \pm0.0088 }  &  \ensuremath{0.194^{+0.030}_{−0.021}} \\
~~~$\log_{10}\langle F \rangle$ (cgs) \tablenotemark{g}
                          \dotfill    &    \ensuremath{9.064\pm0.035}  &  \ensuremath{9.161^{+0.065}_{−0.092}} \\ [-1.5ex]\\
\enddata
\tablenotetext{a}{
    Times are in Barycentric Julian Date calculated directly from UTC {\em without} correction for leap seconds.
    \ensuremath{T_c}: Reference epoch of
    mid transit that minimizes the correlation with the orbital
    period.
    \ensuremath{T_{12}}: total transit duration, time
    between first to last contact;
    \ensuremath{T_{12}=T_{34}}: ingress/egress time, time between first
    and second, or third and fourth contact.
}
\tablecomments{
For HATS-15\,b the fixed-circular-orbit
model has a Bayesian evidence that is $\sim 2$ times larger than the
evidence for the eccentric-orbit model, while for HATS-16\,b the fixed-circular-orbit model has a Bayesian evidence that is $\sim 6$ times larger than the evidence for the eccentric-orbit model. We therefore assume a fixed-circular-orbit in generating the parameters listed here.
}
\tablenotetext{b}{
   Reciprocal of the half duration of the transit used as a jump parameter in our MCMC analysis in place of $\arstar$. It is related to $\arstar$ by the expression $\zrstar = \arstar(2\pi(1+e\sin\omega))/(P\sqrt{1-b^2}\sqrt{1-e^2})$ \citep{bakos:2010:hat11}.
}
\tablenotetext{c}{
    Values for a quadratic law, adopted from the tabulations by
    \cite{claret:2004} according to the spectroscopic (ZASPE) parameters
    listed in \reftabl{stellar}.
}
\tablenotetext{d}{
    For fixed circular orbit models we list
    the 95\% confidence upper limit on the eccentricity determined
    when $\sqrt{e}\cos\omega$ and $\sqrt{e}\sin\omega$ are allowed to
    vary in the fit.
}
\tablenotetext{e}{
    Correlation coefficient between the planetary mass \mpl\ and radius
    \rpl\ estimated from the posterior parameter distribution.
}
\tablenotetext{f}{
    The Safronov number is given by $\Theta = \frac{1}{2}(V_{\rm
    esc}/V_{\rm orb})^2 = (a/\rpl)(\mpl / \mstar )$
    \citep[see][]{hansen:2007}.
}
\tablenotetext{g}{
    Incoming flux per unit surface area, averaged over the orbit.
}
\ifthenelse{\boolean{emulateapj}}{
    \end{deluxetable*}
}{
    \end{deluxetable}
}

\subsection{Rotational modulation}
\label{sec:rota-ctiv}

We inspected the entire light curves obtained from the HATS survey of both HATS-15 and HATS-16, in order to look for possible periodic modulations, caused by the presence of star spots on the surface of the host stars. 

Already with a quick look at the light curves, it is possible to identify a periodicity around 12-13 days for HATS-16, while no perceivable modulation is visible in the HATS-15 photometry.
In order to quantify the variability and check for possible false alarm, we used GLS, a FORTRAN based routine by \cite{zechmeister:2009}, to calculate the generalized Lomb-Scargle periodogram. In brief the data are fitted with a sinusoidal function at different frequencies starting from a provided minimum of 0.0056 [1/d] (that is related to the time range over which the photometric data were acquired) to a maximum frequency of 22.33, with about 79000 steps. A window function is also provided, giving information on the possible false periodicity created by the data sampling. 
We found a peak in the frequencies corresponding to a period of $P=12.350 \pm 0.024$\,days 
and an amplitude of $2.91\pm0.47$\,mmag, with an extremely low false alarm probability \citep[FAP, ][]{cumming:2004}.

As a sanity check, we obtained the periodogram in several different ways always obtaining consistent values of the period. In particular, we calculated the periodogram both for the un-binned and binned data (using several bin sizes) and using also another program \citep[Systemic2, by ][]{meschiari:2009} besides GLS. If the modulation is due to the activity of the star, then the amplitude of the periodic signal may vary in time with the spot coverage, while the period remains constant. We checked for this effect by dividing our data-set into three segments and calculating a periodogram for each one. The three periods are consistent within the error bars, even if in the last dataset the significance of the first peak is lower, as the amplitude of the signal is smaller than that in the first two cases. We assume the variability is due to the apparent spot coverage of the star changing as the star rotates and take the photometric period to be a measure of the stellar rotational period.   

By comparing the rotational period found from the photometry to that inferred from the spectroscopic $v \sin i$ (assuming $\sin i = 1$ and propagating the errors we obtain $P=10.148^{+0.874}_{- 1.102}$\,days), we find a slight inconsistency: the two periods are nearly $2\sigma$ away, and the latter is smaller than the first one. 
However, we stress that the measurement of the $v\sin{i}$ should be considered as an upper limit. It actually describes the broadening of the spectral lines that is caused by a multiplicity of other effects, which are difficult to take into account. Moreover, the rotational period found from photometry doesn't take into account the possibility to have differential rotation. As the latitude of the spot group causing the photometric signal is unknown, the mean period might be larger (if the spots lie around the equator) or smaller (if the spots have a high latitude).

\begin{figure*}[!ht]
\centering
\includegraphics[width=\textwidth]{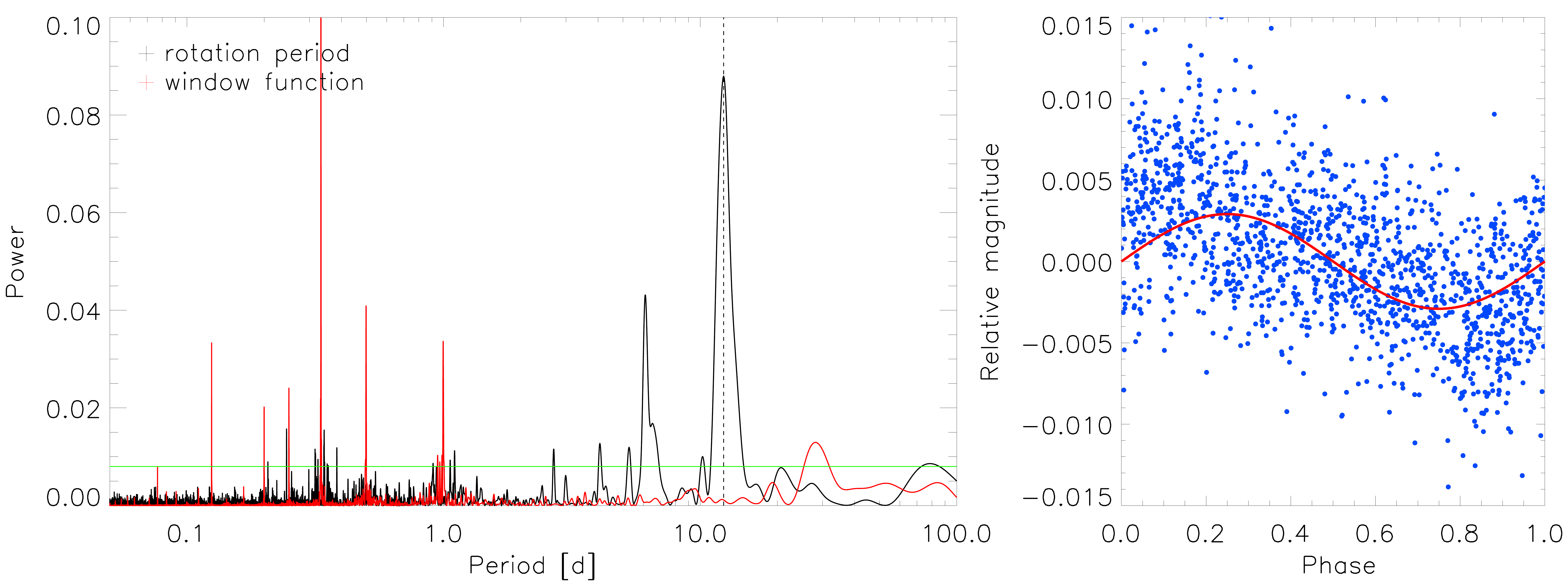}
\caption{\textit{Left:} periodogram for the HATSouth photometric data of HATS-16 obtained during the $\sim$6 months of observations. The horizontal green line represent the false alarm probability FAP= 0.0001, wile the vertical dashed line corresponds to the maximum probability peak. In the same panel the window function is displayed with red lines.  
\textit{Right:} HATSouth photometric data phase folded and binned in 0.001 phase bins for the period (P=12.35 d). The red solid line represents the best sinusoid fit with period fixed to the one found from the periodogram and an amplitude of $\sim$ 2.9\,mmag.}
\label{fig:periodogram}
\end{figure*}

The periodogram, and the phase folded HATSouth photometry with the period of $P=12.35$\,d are shown in \reffigl{periodogram}.

\subsection{Age of the systems}
\label{sec:age}

To determine the age of the two planetary systems, we used different methods. 
A first determination for both the systems was done by modeling the stellar parameters with the Y2 evolutionary model, which was already used in \ref{sec:stelparam}. We found an age of $11.0^{+1.4}_{-2.0}$\,Gyr and $9.5\pm1.8$\,Gyr for HATS-15 and HATS-16, respectively, which suggest that both the systems are old with respect to the average ages of stars in similar spectral class and of exoplanets host in general \citep[e.g.][]{bonfanti:2015}.

Since the rotation of a star is expected to slow down during its lifetime \citep[e.g.][]{soderblom:1983}, it is therefore possible to correlate the stellar period to its age through the gyrochronology. According to \cite{barnes:2007} the errors in the age estimation via gyrochronology for G and K stars are about 28\%.
Following the relations in \cite{barnes:2007}, and using the improved coefficients values from \cite{angus:2015}, we calculated the gyrochronology age of HATS-16 from the rotational period, finding that is younger than 1\,Gyr. Using gyrochronology we also calculated the lower limit for the age of HATS-15 using the period obtained from $v \sin i$ and assuming $i=90\deg$. However the lower limit we found of $\gtrsim 0.5$\,Gyr does not give any useful constraint.  

As expected from the rotational modulation  (\reffigl{periodogram}), the HATS-16 star is slightly active and shows a cromospheric emission in the core of the Ca H \& K absorption lines. We quantified this activity by calculating the activity indices from the ratio of the emission in the line cores, obtaining $S=0.2524$ and $\log{R^{\prime}_{\mathrm{H\,K}}}=-4.644$. Then, using the age-activity relation presented in \cite{mamajek:2008}, we estimated that the expected age of the star is 1.47 Gyr, compatible with the age based on gyrochronology, but substantially younger than the isochrone-based age. 

A further way to constrain the age of a star is to measure its peculiar velocity in the Galactic frame and compare it with those of other stars in our Galaxy, whose age are well determined. Knowing the position of the planetary systems and their kinematics (radial velocity and proper motion measurements from \reftabl{stellar}), one may convert them to peculiar velocities $U$, $V$, $W$ that describe the motion of a star in galactic coordinates. Specifically, $U$ represents the radial component of the velocity, $V$ is the circular component, and $W$ is the vertical component with respect to the Milky Way disk. To compute the peculiar velocity of HATS-15 and HATS-16, we used a web-based calculator\footnote{The calculator can be found at \url{http://www.das.uchile.cl/∼drodrigu/UVWCalc.html}} provided by D. Rodriguez. We then corrected our velocity values for the peculiar motion of the Sun \citep[$U_{\sun}= 7.01 \pm 0.20$\,km\,s$^{-1}$, $V_{\sun}=10.13 \pm 0.12$\,km\,s$^{-1}$ and $W_{\sun}=4.95 \pm 0.09$\,km\,s$^{-1}$,][]{huang:2015}. The results obtained for both the planets are $U=-75.89$\,km\,s$^{-1}$, $V=6.18$\,km\,s$^{-1}$ and $W=-12.03$\,km\,s$^{-1}$ for HATS-15 and $U=-41.72$\,km\,s$^{-1}$, $V=-30.85$\,km\,s$^{-1}$ and $W=-43.00$\,km\,s$^{-1}$ for HATS-16.
For a very young star on a circular orbit, $U$ and $W$ should be close to 0 km\,s$^{-1}$; stars with ages around 1\,Gyr can have a higher dispersion in velocities, with an average value of the order of $10-12$\,km\,s$^{-1}$; finally, at an age around 10\,Gyr, the velocities are expected to be of the order of $25-35$\,km\,s$^{-1}$ \citep{binney:2000}.
This increase in velocity dispersion with age can be explained taking into account the fact that stars, initially formed in the galactic plane (where molecular clouds and star forming region are mainly located), interact with the galactic environment and gain vertical and radial velocity component in the galactic reference frame \citep[][and reference therein]{nordstrom:2004}. The peculiar velocities of our two stars, compared with the average velocities for the stars in our Galaxy, point towards an old age, which is in good agreement with the ones that we found from the isochrone fitting.

\subsection{Excluding blend scenarios}
\label{sec:blend}

One of the most common false positive scenarios for candidates produced by transiting surveys are blends. We attempt to exclude the possibility that our two planetary systems are not hosting planets, but instead are diluted eclipsing binaries, following \citet{hartman:2012:hat39hat41}.

We modeled the available photometric data for each object as a blend between an eclipsing binary star system and a third star along the line of sight. The physical properties of the stars were constrained using the Padova isochrones \citep{bertelli:2008}, while we also required that the brightest of the three stars in the blend had atmospheric parameters consistent with those measured with ZASPE. 

We found that for both HATS-15 and HATS-16, a model consisting of a single star with a transiting planet provides a lower $\chi^2$ fit to the available photometric data than any of the blended stellar eclipsing binary models tested. Based solely on the photometry, for HATS-15 we ruled out blend models with $\sim 1\sigma$ confidence, while for HATS-16 we rule them out with $\sim 2\sigma$ confidence. Moreover, we found that for both systems any blend model that could plausibly fit the photometry (i.e., which cannot be rejected with greater than $5\sigma$ confidence) would have been easily identified as a composite system based on the spectroscopic observations. 
Indeed, we simulated the cross-correlation functions for these possible blend systems, finding that in all the cases, at some of the observed phases, a double peak should have been seen, and that all of the
blend scenarios which could plausibly fit the photometry of either system would have produced large RV and BS variations (greater than 1\,$\mathrm{km\,s}^{-1}$), which is in conflict with the observations.
We conclude, therefore, that both HATS-15 and HATS-16 are transiting planet systems, and that neither object is a blended stellar eclipsing binary system.

Albeit we can rule out the possibility that the two systems are not blended by an eclipsing binary, we cannot rule out the possibility that one of the two systems is a transiting planet system, whose photometry is diluted by the light coming from an unresolved stellar component. 
For HATS-15 we find that a system consisting of a $0.84\,\msun$ star with a transiting planet and a $0.65$\,\msun\ binary companion provides a slightly better fit to the photometric data ($\sim 1.5\sigma$) than the best-fit single-star-with-planet model. The secondary-to-primary $V$-band light ratio of this binary system is 14\%, which would be marginally detectable in the observed CCFs unless the system were near conjunction. We also find that a binary star companion of any mass, up to that of the transiting planet host, provides a fit to the photometric data which cannot be distinguished from the best-fit single-star-with-planet model. For HATS-16 we find that binary star systems with a secondary mass between $0.65\,\msun$ and $0.8\,\msun$ can be ruled out at $3\sigma$ confidence based on the photometry. This is driven by the broad-band photometric colors which are consistent with the measured effective temperature of the primary star assuming no reddening. Binary companions close in mass to the primary yield similar photometric colors and are not ruled out, while
binary companions below $0.65$\,\msun\ are too faint to significantly affect the photometric colors.  Higher spatial resolution imaging and/or continued RV monitoring would be needed to search for binary star companions to either system. For the remainder of the paper we assume that each of the two systems is an isolated star with a close-in transiting planet.

\subsection{Global modeling of the data}
\label{sec:globmod}

In order to measure the physical properties of the two planets, we modeled all the photometric and spectroscopic data in our possession following the same approach as in \citet{pal:2008:hat7,bakos:2010:hat11,hartman:2012:hat39hat41}. 

As far as it concerns the light curves, both the HATSouth and the follow-up ones, the fit was performed using the transit models from \citet{mandel:2002}, employing a quadratic law to describe the limb darkening effect, and fixing the coefficients to those from \cite{claret:2004}. In the case of the HATSouth photometry, in modelling the transit depth, we included an extra parameter describing the possible dilution caused by the blending of neighboring stars, and the over-correction by the trend-filtering method. To correct for systematic errors in the photometry of the follow-up light curves, a quadratic trend was included in the model of each transit event. 
The RV data were fitted with Keplerian orbits, allowing the zero-point and the RV jitter for each instrument to vary independently as a free parameter.

To determine the posterior distribution of the parameters and obtain the relative uncertainties we used a Differential Evolution Markov Chain Monte Carlo \citep[DEMCMC][]{terbraak:2006,eastman:2013}. 

We fitted both fixed circular orbits and free-eccentricity models to the data, and for both systems found that the data are consistent with a circular orbit. As for both systems the fixed circular orbit model had a higher Bayesian evidence, we adopted the parameters assuming no eccentricity for either object.

The parameters obtained from this analysis for each system are listed in \reftabl{planetparam}. In brief we found that HATS-15b has a mass of $M =2.17\pm0.15\, M_{\mathrm{J}}$ and a radius of  $R =1.105\pm0.040\, R_{\mathrm{J}}$, resulting in a bulk density of $\rho =1.48\pm0.18\, \rho_{\mathrm{J}}$. For HATS-16b, we found that it is more massive, $M = 3.27\pm0.19\, M_{\mathrm{J}}$, and larger, $R =1.30\pm0.15\, R_{\mathrm{J}}$, which imply a slightly lower density, $\rho =1.39_{-36}^{+0.71}\, \rho_{\mathrm{J}}$.

\section{Discussion and Conclusion}
\label{sec:discussion}

\begin{figure*}
\centering
\includegraphics[width=18.0cm]{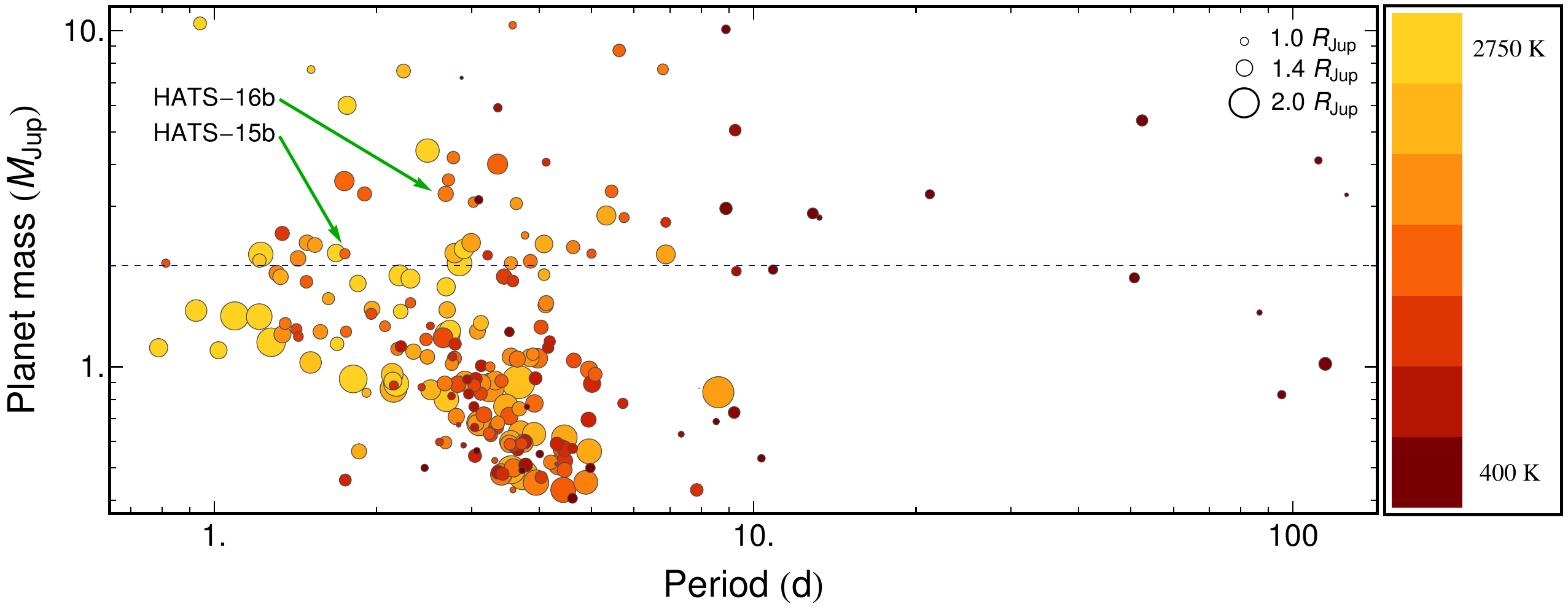}
\caption{Mass--period diagram of transiting exoplanets in the mass range $0.4\,M_{\mathrm{J}}<M_{\mathrm{p}}<10\,M_{\mathrm{J}}$. The
planets are represented by circles, whose size is proportional to planet radius. Color indicates equilibrium temperature. The dashed line demarcates the high-mass regime ($M_{\mathrm{p}}>2\,M_{\mathrm{J}}$). The error bars have been suppressed for clarity. Data taken from the Transiting Extrasolar Planet Catalogue (TEPCat), which is available at http://www.astro.keele.ac.uk/jkt/tepcat/.}
\label{fig:P-m2}
\end{figure*}

We have presented the discovery of HATS-15b and HATS-16b, two massive hot Jupiters orbiting around old dwarf stars. By carefully analyzing the photometric and spectroscopic data, we can exclude false positive scenarios, confirming the planetary nature of the transiting bodies. We found that HATS-15 is a planetary system, $\sim 690$\,pc far away from our Sun, that hosts a $2.17\pm0.15\, M_{\mathrm{J}}$ hot Jupiter, which orbits around a G9\,V star in $\sim 1.75$\,days. HATS-16b is a massive $3.27\pm0.19\, M_{\mathrm{J}}$ hot Jupiter $\sim 770$\,pc away from us, and is orbiting a G3\,V star in $\sim 2.69$\,days. \reffig{P-m2} shows the locations of the two new planets in the period--mass diagram, together with the other known transiting exoplanets. 

In \refsec{age}, we have presented several methods to estimate the stellar ages. 
In the case of HATS-15, the two different methods that we used (Y2 isochrones fitting and stellar kinematics) gave consistent values; we therefore adopted the value found from the modeling with the Y2, which dates the star to be $11.0^{+1.4}_{-2.0}$\,Gyr old. 

Concerning HATS-16, instead, we found a strong discrepancy between the age measured with Y2 and stellar kinematics, compared with the one obtained with gyrochronology and stellar activity. However, we believe that the age obtained from the stellar evolutionary models ($9.5\pm1.8$\,Gyr) is more reliable than the very young age found based on its activity. Indeed, an old age is more compatible with the slightly low density of the star, compared to what is expected for a young star with its effective temperature and metallicity.
By studying all the exoplanetary host stars for which the rotational period is measured, \cite{maxted:2015} found that there is no evidence for the gyrochronological ages to be on average smaller than those measured from isochronal models. Therefore the cause of the discrepancy seen in our measurements has to be searched elsewhere, and not in a measurement bias.
The short rotation period (12.4\,d) of HATS-16, may be explained in a planet-star interaction context, in which the star has been tidally spun up by the planet \citep[e.g.][]{pont:2009, husnoo:2012}. Given the mass of the planet, its distance from the star, and assuming a stellar tidal quality factor (i.e. the efficiency of tidal dissipation in the star) $Q_{*}$=10$^{6}$, obtained from a rough estimation of the orbital evolution timescale \citep[following][]{penev:2011}, we calculated that just few Gyr are necessary for spinning up the star to its current rotation (here we have assumed that all the angular momentum is dumped in the convective zone, and that only a few percent change in the orbital period is needed).

Considering the entire population of known gas planets, most of the planets in the high-mass regime are found to orbit their star at a large distance, and just few tens are hot Jupiters (see \reffig{P-m2}). 
The rareness of massive giants with $M\ga2\,M_{\mathrm{J}}$ and $P\la5$\,d cannot be explained by observational biases, since the two more efficient detection methods (transit and RV) are more prone to find massive planets in short orbits than in larger ones \footnote{In some cases, massive close-in giant planets may spin-up their parent star, causing higher values of its $v\sin{i}$ than expected, as in the present case of HATS-16 ($v \sin i=6.17$). Then, a possible planetary candidate identified by a transiting or RV survey may be discarded, and not be further followed up, because of its fast rotating host star \citep{pont:2009}. For this reason, planet-candidate rejections based on $v\sin{i}$ should be relaxed in order to do not miss possible massive hot Jupiters.}

\begin{figure}[!ht]
\centering
\includegraphics[width=\columnwidth]{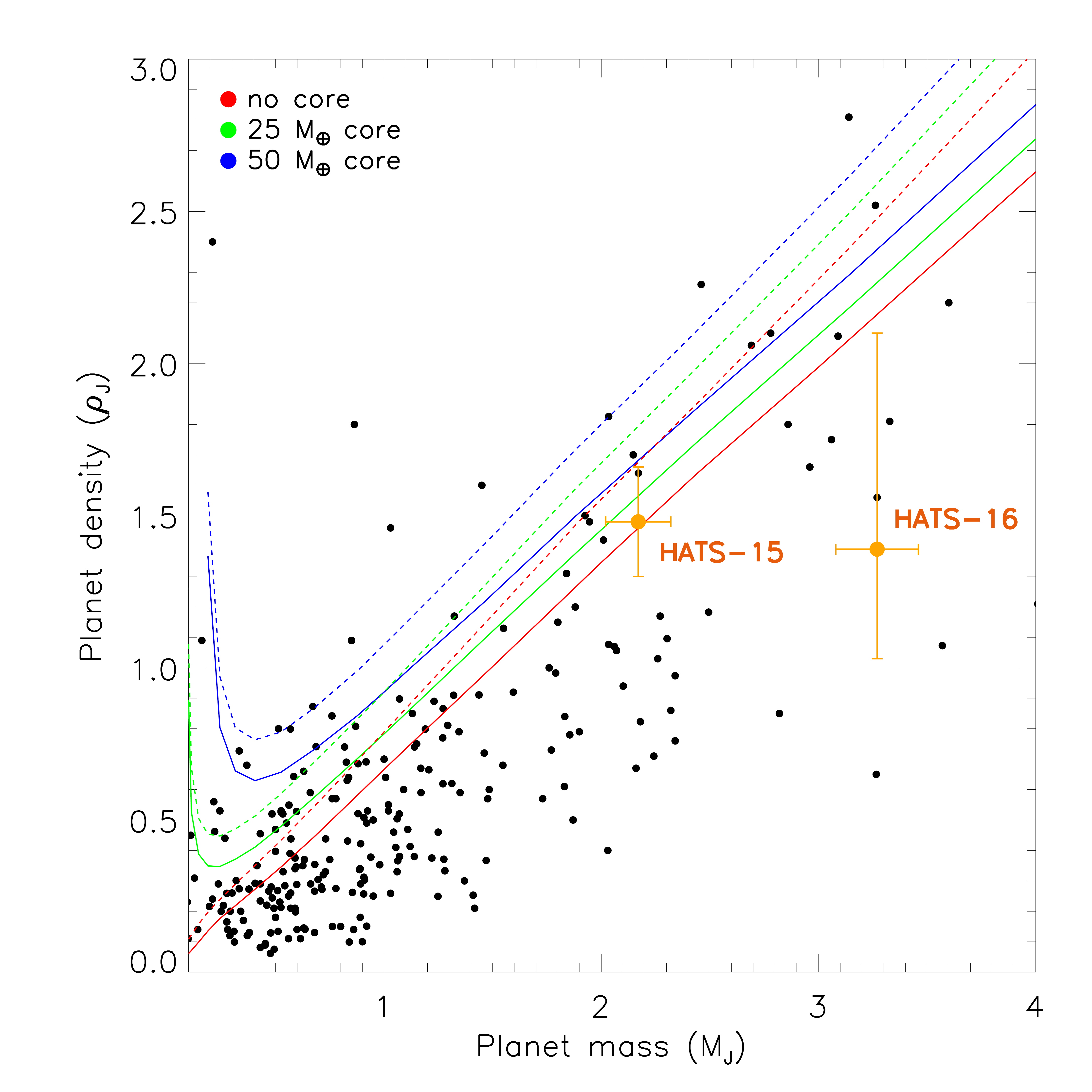}
\caption{The HATS-15 and HATS-16 planets (highlighted in orange) are presented in the mass--density diagram of the known transiting planets for which the mass is measured. The superimposed lines represent the expected density of the planet having an inner core of 0, 25 and 50 Earth masses, and calculated for 10 Gyr old planets at 0.02 AU, solid lines, and 0.045 AU, dashed lines \citep{fortney:2007}. To be noticed is that $\sim$2\,-\,3\,$M_{\mathrm{J}}$ transiting planets are much rarer than those in the $\sim$0.1\,-\,1\,$M_{\mathrm{J}}$ mass regime.}
\label{fig:m-rho}
\end{figure}

The reason of this paucity should be related to different channels that were undertaken by these planets during their formation and migration history. Alternatively, they are simply rarer than the lighter hot Jupiters, as it happens between planets and brown dwarfs. Actually, the existence of the brown-dwarf desert (i.e. companions in the mass range $10-100\,M_{\mathrm{J}}$ are $\sim 1$ order-of-magnitude rarer compared to less massive objects, e.g., \citealp{winn:2014} and references therein) is by now well ascertained. Finding and characterizing new hot Jupiters is the only key to clarify the causes of the rareness of massive hot Jupiters and the ongoing ground-based surveys, as well as the upcoming \citep[TESS,][]{ricker:2009} and future \citep[PLATO,][]{catala:2006} space missions, are essential for this purpose.

Using the planetary mass--density relations for planets with $M_{\mathrm{p}} \gtrsim 0.4\, M_{\mathrm{J}}$ presented in \cite{bakos:2015:hats7}, we find that both the planets fall within the predicted range, in particular HATS-15b shows an average density among the planets with the same mass, while HATS-16b lies close to the lower limit of the relation.   
In addition, comparing the mass and radius of the two planets with the predictions of \cite{fortney:2007}, we give some constraints on the presence of an inner heavy core. According to the distance from its parent star, the age of the system and for a fixed planetary mass, the expected radius of the planet is computed in the cases of different internal core masses. 
We found that HATS-16b is well described by a core-less model, whereas in the case of HATS-15b we expect that it has a very light core. Indeed, we can exclude the case of a massive core as our measurement is only consistent with models predicting a core-mass lower than $50\,M_{\oplus}$ (see \reffig{m-rho}).

Comparing HATS-16\,b with known planets in the same mass range, we found that the majority of the planets are located far away from their host star having periods greater than 200 days. The eccentricity distribution for those planets seems to be more flat than for the confirmed planets with smaller masses, where there is a slight preference for circular orbits. Howbeit, this can be easily explained taking into account the fact most of those planets have very short periods and therefore are more prone to suffer from circularizations and tidal locking mechanism that result in circular orbits.

In the mass--period parameter space the twin systems to HATS-16 are Kepler-43 \citep{bonomo:2012} and WASP-10 \citep{christian:2009}. The first system doesn't show any hint of stellar activity or planet-star interaction, however this might not be surprising given the young age of the system. In the latter case the planet has an eccentric orbit that, assuming a $Q_{*}\sim10^{5}$ or $\sim10^{6}$, should have disappeared because of circularizations due to tidal interaction, but still persist indicating some other form of interaction e.g. the presence of an external massive body belonging to the same system.

Although having a mass twice that of Jupiter, HATS-15\,b lies in a more populated region of the parameter space than HATS-16\,b. There are indeed both long period and short period planets in its mass range that share their average properties with all the other planets. HATS-15\,b finds its twin planet in WASP-87A\,b \citep{anderson:2014:wasp87}. The main difference between the two planetary systems is that while HATS-15 is a single star, WASP-87 is composed by two stars: the mid-F type host star and a mid-G companion. More over given the different spectral class of the host star HATS-15\,b has a less inflated radius than its twin and is therefore denser.

\acknowledgements 

Development of the HATSouth project was funded by NSF MRI grant NSF/AST-0723074, operations have been supported by NASA grants NNX09AB29G and NNX12AH91H, and follow-up observations receive partial support from grant NSF/AST-1108686.
A.J.\ acknowledges support from FONDECYT project 1130857, BASAL CATA PFB-06, and project IC120009 ``Millennium Institute of Astrophysics (MAS)'' of the Millenium Science Initiative, Chilean Ministry of Economy. R.B.\ and N.E.\ are supported by CONICYT-PCHA/Doctorado Nacional. R.B.\ and N.E.\ acknowledge additional support from project IC120009 ``Millenium Institute of Astrophysics  (MAS)'' of the Millennium Science Initiative, Chilean Ministry of Economy.  V.S.\ acknowledges support form BASAL CATA PFB-06.  M.R.\ acknowledges support from FONDECYT postdoctoral fellowship 3120097.
This work is based on observations made with Telescopes at the ESO La Silla Observatory.
This paper also uses observations obtained with facilities of the Las Cumbres Observatory Global Telescope.
Work at the Australian National University is supported by ARC Laureate Fellowship Grant FL0992131.
We acknowledge the use of the AAVSO Photometric All-Sky Survey (APASS), funded by the Robert Martin Ayers Sciences Fund, and the SIMBAD database, operated at CDS, Strasbourg, France.
The imaging system GROND has been built by the high-energy group of MPE in collaboration with the LSW Tautenburg and ESO\@.  

We thank F. Rodler, W. Brandner, M. W\"{o}llert and J. Schliederfor useful comments and advices.  We thank Helmut Steinle and Jochen Greiner for supporting the GROND observations presented in this manuscript.
We are grateful to P.Sackett for her help in the early phase of the HATSouth project.

\bibliographystyle{apj}
\bibliography{hatsbib}

\clearpage

\tabletypesize{\scriptsize}
\ifthenelse{\boolean{emulateapj}}{
    \begin{deluxetable*}{lrrrrrl}
}{
    \begin{deluxetable}{lrrrrrl}
}
\tablewidth{0pc}
\tablecaption{
    Relative radial velocities and bisector spans for HATS-15 and HATS-16.
    \label{tab:rvs}
}
\tablehead{
    \colhead{BJD} &
    \colhead{RV\tablenotemark{a}} &
    \colhead{\ensuremath{\sigma_{\rm RV}}\tablenotemark{b}} &
    \colhead{BS} &
    \colhead{\ensuremath{\sigma_{\rm BS}}} &
    \colhead{Phase} &
    \colhead{Instrument}\\
    \colhead{\hbox{(2,456,000$+$)}} &
    \colhead{($\mathrm{m\,s}^{-1}$)} &
    \colhead{($\mathrm{m\,s}^{-1}$)} &
    \colhead{($\mathrm{m\,s}^{-1}$)} &
    \colhead{($\mathrm{m\,s}^{-1}$)} &
    \colhead{} &
    \colhead{}
}
\startdata
\multicolumn{7}{c}{\bf HATS-15} \\
\hline\\
$ -188.27661 $ & $   418.77 $ & $    54.00 $ & $  -75.0 $ & $   23.0 $ & $   0.677 $ & FEROS \\
$ -186.41173 $ & $   418.77 $ & $    32.00 $ & $   16.0 $ & $   15.0 $ & $   0.744 $ & FEROS \\
$ -185.23774 $ & $  -279.23 $ & $    45.00 $ & $    0.0 $ & $   19.0 $ & $   0.416 $ & FEROS \\
$ 81.70953 $ & $  -270.39 $ & $   131.00 $ & $ -345.0 $ & $   37.0 $ & $   0.176 $ & Coralie \\
$ 82.73890 $ & $   769.61 $ & $   189.00 $ & $ -124.0 $ & $   37.0 $ & $   0.766 $ & Coralie \\
$ 146.71065 $ & $  -344.23 $ & $    75.00 $ & $ -287.0 $ & $   30.0 $ & $   0.373 $ & FEROS \\
$ 162.62586 $ & $     9.77 $ & $    49.00 $ & $  107.0 $ & $   21.0 $ & $   0.481 $ & FEROS \\
$ 167.63696 $ & $  -337.23 $ & $    32.00 $ & $   57.0 $ & $   15.0 $ & $   0.348 $ & FEROS \\
$ 172.74493 $ & $  -468.23 $ & $    34.00 $ & $   27.0 $ & $   15.0 $ & $   0.271 $ & FEROS \\
$ 241.56988 $ & $   212.61 $ & $   126.00 $ & $  -48.0 $ & $   37.0 $ & $   0.657 $ & Coralie \\
$ 400.84394 $ & $   389.77 $ & $    35.00 $ & $   44.0 $ & $   16.0 $ & $   0.801 $ & FEROS \\
$ 404.85903 $ & $  -189.23 $ & $    39.00 $ & $   75.0 $ & $   17.0 $ & $   0.099 $ & FEROS \\
$ 406.82914 $ & $  -261.23 $ & $    43.00 $ & $   71.0 $ & $   19.0 $ & $   0.226 $ & FEROS \\
$ 424.80863 $ & $    48.77 $ & $    41.00 $ & $   10.0 $ & $   18.0 $ & $   0.515 $ & FEROS \\
$ 426.78243 $ & $   138.77 $ & $    85.00 $ & $   10.0 $ & $   34.0 $ & $   0.644 $ & FEROS \\
$ 427.78530 $ & $  -342.23 $ & $    84.00 $ & \nodata      & \nodata      & $   0.218 $ & FEROS \\
\hline\\
\multicolumn{7}{c}{\bf HATS-16} \\
\hline\\
$ 841.86396 $ & $  -477.43 $ & $    18.00 $ & $   63.0 $ & $   16.0 $ & $   0.355 $ & FEROS \\
$ 844.89323 $ & $   -54.43 $ & $    16.00 $ & $   43.0 $ & $   14.0 $ & $   0.483 $ & FEROS \\
$ 846.85037 $ & $  -465.43 $ & $    14.00 $ & $   44.0 $ & $   13.0 $ & $   0.211 $ & FEROS \\
$ 852.89702 $\tablenotemark{c} & $  -435.43$ & $   19.0$ & $   -215 $ & $   16.0$ & $ 0.462 $ & FEROS \\
$ 853.84870 $\tablenotemark{c} & $   439.57$ & $   23.0$ & $   -363 $ & $   19.0$ & $ 0.816 $ & FEROS \\
$ 854.90341 $\tablenotemark{c} & $  -482.43$ & $   19.0$ & $   -429 $ & $   16.0$ & $ 0.208 $ & FEROS \\
$ 858.68871 $ & $   292.57 $ & $    15.00 $ & $  -36.0 $ & $   14.0 $ & $   0.618 $ & FEROS \\
$ 858.86912 $ & $   500.57 $ & $    16.00 $ & $   23.0 $ & $   14.0 $ & $   0.685 $ & FEROS \\
$ 866.80486 $ & $   342.57 $ & $    17.00 $ & $   69.0 $ & $   15.0 $ & $   0.639 $ & FEROS \\
$ 871.70647 $ & $  -133.43 $ & $    17.00 $ & $   56.0 $ & $   16.0 $ & $   0.464 $ & FEROS \\
$ 908.12221 $ & $   -57.08 $ & $     5.94 $ & \nodata      & \nodata      & $   0.019 $ & HIRES \\
$ 909.06821 $ & $  -290.30 $ & $     9.17 $ & \nodata      & \nodata      & $   0.371 $ & HIRES \\
$ 910.05021 $ & $   477.26 $ & $     6.37 $ & \nodata      & \nodata      & $   0.736 $ & HIRES \\
$ 912.07720 $ & $   -94.72 $ & $     6.24 $ & \nodata      & \nodata      & $   0.491 $ & HIRES \\
$ 932.74739 $ & $  -375.43 $ & $    16.00 $ & $   48.0 $ & $   15.0 $ & $   0.185 $ & FEROS \\
$ 942.64274 $ & $   429.57 $ & $    18.00 $ & $   39.0 $ & $   16.0 $ & $   0.868 $ & FEROS \\   
\enddata
\tablenotetext{a}{
    The zero-point of these velocities is arbitrary. An overall offset
    $\gamma_{\rm rel}$ fitted independently to the velocities from
    each instrument has been subtracted.
}
\tablenotetext{b}{
    Internal errors excluding the component of astrophysical jitter
    considered in \refsecl{globmod}.
}
\tablenotetext{c}{
    These observations were excluded from the analysis because the
    extracted spectra had significant contamination from scattered
    moonlight leading to large systematic errors in the measured RVs and BSs.
}
\ifthenelse{\boolean{rvtablelong}}{
}{
} 
\ifthenelse{\boolean{emulateapj}}{
    \end{deluxetable*}
}{
    \end{deluxetable}
}

\end{document}